\documentclass[%
twocolumn,
 amsmath,amssymb,
 aps,
prb,
]{revtex4-2}

\pdfoutput=1

\usepackage{comment}
\usepackage{lipsum}

\usepackage{hyperref}

\usepackage{graphicx}% Include figure files
\usepackage{dcolumn}% Align table columns on decimal point
\usepackage{bm}% bold math
\usepackage[english]{babel}
\usepackage{float}
\usepackage{natbib}
\begin{document}

\preprint{APS/123-QED}

\title{Quasidisorder Induced Topology}

\author{M. F. Madeira}
\affiliation{Departamento de F\'{i}sica and CeFEMA, Instituto Superior T\'{e}cnico, Universidade de Lisboa, Av. Rovisco Pais, 1049-001 Lisboa, Portugal}
\author{P. D. Sacramento}
\affiliation{Departamento de F\'{i}sica and CeFEMA, Instituto Superior T\'{e}cnico, Universidade de Lisboa, Av. Rovisco Pais, 1049-001 Lisboa, Portugal}
\date{\today}

\begin{abstract}
We study the effects of quasidisorder and Anderson disorder on a two dimensional topological 
superconductor with an applied external magnetic field. The cases of a $p$-wave superconductor and a noncentrosymmetric superconductor with mixed $p$ and $s$-wave pairings and Rashba spin-orbit coupling are studied.
We show that, for a perpendicular magnetic field, the introduction of quasidisorder leads to the appearance of topological phases in new regions, characterised by an integer value of the Chern number.
For a parallel magnetic field, we identify regimes with the appearance of new Majorana flat bands and also new unidirectional Majorana edge states, as quasidisorder is introduced. We show that the Majorana flat bands have a quantized Berry phase of $\pi$ and identify it as a topological invariant. Two topological transitions are identified and  the values of the critical exponents $z$ and $\nu$ are obtained. The fractal nature of the eigenstates is discussed both for Anderson disorder and Aubry-Andr\'e disorder.
\end{abstract}

%\keywords{Suggested keywords}%Use showkeys class option if keyword
                              %display desired
\maketitle

%\tableofcontents

\section{\label{sec:intro}Introduction}
The search and study of topological properties of matter has proved fruitful in recent years in research in materials science and condensed matter physics. Superconductors have long been a focus of interest due to their promising applications. Superconductors with intrinsic topological properties, in particular, have recently attracted theoretical and experimental interest due to phenomena associated with surface or edge Majorana modes, which appear from an interplay between topology and bulk-boundary correspondence \cite{beenakker_search_2013, hasan_topological_2010, qi_topological_2011}. These Majorana zero modes emerge with non-Abelian exchange statistics and are sought after due to their promising expected applications in quantum computing, being candidates for the building blocks of a quantum qubit \cite{alicea_2012, sato_ando_2017}.
\par
It has been theoretically predicted that Majorana states appear as flat dispersion bands in gapless superconducting phases, such as in the $d_{xy}$+$p$-wave pairing noncentrosymmetric superconductor in two dimensions with preserved time-reversal symmetry \cite{tanaka_anomalous_2010,sato_topology_2011}, or for a $p$-wave topological superconductor in two dimensions, with broken time reversal symmetry by an applied magnetic field parallel to the two dimensional plane of the system \cite{patricklee}. Flat bands also emerge on the surface of three dimensional noncentrosymmetric superconductors, with spin-orbit coupling and which preserve time-reversal symmetry \cite{schnyder_topological_2011, schnyder_types_2012}. 
%\textbf{Kagome} 
It is predicted that flat bands can increase the critical temperature for superconductivity, and even give rise to room-temperature superconductivity \cite{Volovik,torma}. 
Similar behavior has been found when one has finite-size systems (with increased fluctuations of
the density of states) \cite{sangita,pnictides}, 
non-homogeneous order parameters 
\cite{burmistrov1,mayoh0,bofan2,carbillet,lebarski,lebarski2},
or fractal (critical) states 
\cite{feigelman1,feigelman2,mayoh1,zhao,verdu,bofan1,fan,stosiek,zhang}
with corresponding
spatial fluctuations of the amplitude of the wave functions.
The difference between an isolated flat band and a flat band with band touchings has also been recently discussed \cite{arxivpaper}. It was shown that isolated flat bands are not needed to achieve a higher superconducting temperature, and that band touchings can actually increase it. Flat electronic bands can also be found in some Kagome-type superconductors \cite{kagome1}. A growing interest has been seen in these types of materials, AV$_3$Sb$_5$ (with A$=$K, Rb, Cs), which can host exotic quantum properties, displaying topological phases, an unconventional charge density wave, and evidence of time-reversal symmetry breaking \cite{kagome2,kagome3,kagome4,kagome5}.

\par
The study of perturbations in condensed matter systems, namely through the introduction of disorder, is a central issue. On one hand, introducing disorder can destroy some phases and their properties, preventing their experimental observation. In this sense, the study of their robustness becomes crucial. On the other hand, disorder can by itself lead to new phenomena or stabilize previously existing phases.
One type of disorder that has been attracting interest in the research field is quasiperiodic disorder. These systems are somewhat in between periodic and truly random systems, and exhibit interesting phenomena, in transport \cite{Sutradhar,saha}, topological properties \cite{lang_edge_2012,tezuka_reentrant_2012,degottardi_majorana_2013,cai_topological_2013,kitaevaa,cai_quantum_2014,PhysRevLett109106402,nakajima_competition_2021,longhi_topological_2020,zilberberg_topology_2021,liu_topological_2018,rosa_exploring_2021,verbin_observation_2013,kraus_topological_2012}, and critical behaviour \cite{Sutradhar,pixley_wilson_huse_gopalakrishnan_2018,goblot_gratiet_harouri_sagnes_ravets_etal_2020,xiao_observation_2021}. It is possible to realize these types of systems in experimental setups of ultracold atoms \cite{an_engineering_2018, roati_anderson_2008}, in optical lattices  \cite{opticallattices, luschen_single_particle_2018} or in photonics systems  \cite{lahini_observation_2009}. In addition to systems subject to quasiperiodic potentials, as in the Aubry-Andr\'e model \cite{aubry_andre_1980}, there has been growing interest in Moir\'e systems in which two incommensurate lattices are connected, or in which layers of lattices are put in contact and rotated, such as the 2d twisted bilayer graphene \cite{lopesdossantos_graphene_2007,bistritzer_moire_2011,lopes_dos_santos_continuum_2012,mao_senthil_2021,goncalves_olyaei_amorim_mondaini_ribeiro_castro_2021}. In such systems, a superlattice potential is created from proximity coupling between the two lattices, which, depending on the angle of rotation between the two, may exhibit quasiperiodicity.

\par An example of the study of coexisting quasidisorder and superconductivity, which is significant in the context of this work, is the one dimensional Kitaev chain with Aubry-Andr\'e modulation \cite{wang_liu,fraxanet_bhattacharya_grass_rakshit_lewenstein_dauphin_2021,tong_meng_jiang_lee_neto_xianlong_2021,lv_quantum_2022}. 
Without superconductivity the model has a topological nature revealed by its mapping to a $2d$ quantum Hall
system \cite{hofstadter}, maintaining a topological nature as we add superconducting pairing. In general,
the mappings involve a corresponding model in a higher dimension in the form of some
parent Hofstadter generalized Hamiltonian.
Topology in quasicrystals may be understood considering mappings to higher dimensions, typically
of the types
$1d$ to $2d$ and $2d$ to $4d$.
In $1d$ with no superconductivity the model
is self-dual (position and momentum space) and there is a single transition from an extended state phase
to a phase where all the states are localized. At the transition point the system has critical states.
Generalized models show the existence of mobility edges, such that there is a separation as a function
of energy between extended and localized states
\cite{dassarma1,dassarma2,dassarma3,ganeshan,gopola,logan,liu,ganesham2} and the existence
of hidden dualities leads to a rich class of systems, where such edges appear \cite{hidden}.
The introduction of $p$-wave pairing in the Aubry-Andr\'e model leads to the appearance of a  
finite extent region of critical (fractal) states, between the regions of extended and localized states.
Remarkably, the transitions between localised and critical regimes have been studied and were found to deviate from the known Aubry-Andr\'e universality class \cite{tong_meng_jiang_lee_neto_xianlong_2021,lv_quantum_2022}.

\par In this work we study a model of a two-dimensional superconductor with spin triplet $p$-wave pairing, or mixed $p$ and $s$-wave pairings with Rashba spin-orbit coupling, in the presence of a time reversal symmetry breaking magnetic field. Some materials which are candidates for realizing triplet pairing superconductivity include Sr$_2$RuO$_4$ \cite{Sr2RuO4}, UPt$_3$ \cite{UPt3} and  Cu$_x$Bi$_2$Se$_3$ \cite{CuxBi2Se3}. In the presence of $s$-wave pairing and Rashba spin orbit coupling, the model describes a noncentrosymmetric superconductor, of which are examples $\text{CePt}_{3}\text{Si}$ \cite{bauer_heavy_2004}, $\text{CeIrSi}_{3}$ \cite{sugitani_pressure-induced_2006} and $\text{CeRhSi}_{3}$ \cite{kimura_pressure-induced_2005}. In the noncentrosymmetric regime the breaking of inversion symmetry allows for the mixture of spin-triplet and spin-singlet pairings. This mixing is expected to lead to novel phenomena such as higher than usual values of the upper critical field \cite{fujimoto_electron_2007, frigeri_superconductivity_2004}.
\par The clean model has been studied, in both the centrosymmetric and the noncentrosymmetric regimes, and is known to possess diverse topological properties. If time-reversal symmetry is preserved, the model displays gapless Majorana edge states and is characterised by a $\mathbf{Z}_{2}$ invariant. The observed properties when time-reversal symmetry is broken by an external magnetic field are found to be very dependent on its direction in relation to the two-dimensional superconducting plane. If the magnetic field is such that it is perpendicular to the plane of the superconductor, the model has a rich phase diagram indexed by the Chern number \cite{sato_fujimoto_2009}. When the magnetic field is parallel to the plane of the system, interesting phenomena, such as Majorana flat bands or Majorana unidirectional states, appear on phases with a gapless bulk \cite{patricklee,yanase1}. 

The effect of disorder may be considered in different ways. One possibility is to consider a non-homogeneous magnetic field, achieved by inserting magnetic impurities in the clean superconductor \cite{balatsky} which may give rise to or change topological properties in the system. Examples include the addition of chains of magnetic adatoms \cite{Reis,Sacramento}, islands of magnetic impurities \cite{Ojanen} or fully random distributions of impurities \cite{sacramento_cadevz_mondaini_castro_2019}. Another possibility is to consider potential scattering impurities on the superconductor in the presence of a constant magnetic field, either perpendicular or parallel to the system. We are interested in studying the effects of quasidisorder in these regimes. Besides Aubry-Andr\'e disorder, we will also consider Anderson disorder as a comparison to the effects of quasi-periodicity.

Anderson localization does not require full randomness. If differences in potential between
sites are large enough compared to hoppings, one may expect a transition to localized states.
In addition to full randomness, a quasidisordered potential leads to localization if the disorder
amplitude is large enough \cite{devakul}.
One expects that the Aubry-Andr\'e quasiperiodic potential should affect the long-range nature of states,
and in particular topological states that are by themselves of long-range nature.
Aubry-Andr\'e is expected to be naturally of a multifractal nature.
Anderson and Aubry-Andr\'e are different and, in particular, critical states due to Anderson appear at
the transition to localization while in the Aubry-Andr\'e added to the Kitaev $1d$ model 
one finds phases with this behavior
(or in $2d$ a mixture of critical states in the crossover to localization).
Multifractality probes long distances and therefore one expects that it
may enhance superconductivity due to Chalker scaling \cite{chalker1,chalker2}, as expected and
observed with other inhomogeneities.
Multifractal wave functions have larger spatial overlap and stronger state to state correlations for states
with similar energies.

As stated previously, quasiperiodicity may also lead to topological properties \cite{huang,pixley}.
A $2d$ topological insulator plus quasiperiodic potential shows a transition from a trivial insulator 
to a topological insulator.
Flat topological bands and eigenstate criticality have also been shown as a result of a quasiperiodic perturbation
in the context of the Bernevig-Hughes-Zhang model plus
$2d$ quasiperiodic potential \cite{fu}.

The presence of gapless states in a system may also be associated with long-distance behavior. For instance,
nodal points of Weyl semimetals may lead to interesting behavior in the presence of disorder. 
It has been shown that they 
survive the presence of moderate disorder \cite{altland}.
On the other hand, in the case of gapless states of the form of nodal loops, any amount of disorder mixes states.
Disorder-driven multifractality has been shown in Weyl nodal loops \cite{nodalloop}.
In the case of magic angle semimetals quasiperiodicity generically leads to flat bands in nodal, 
semi-metallic structrures.
A transition from a Weyl semimetal to metal driven by quasiperiodic potential has been found in $3d$
\cite{pixley_wilson_huse_gopalakrishnan_2018,fu2}.

While the influence of disorder, either Anderson or quasidisorder has been extensively
considered in the case of one-dimensional systems, including in the presence of
superconductivity, it is interesting to consider their
effects on a two-dimensional $p$-wave superconductor, and in particular in the presence of a
magnetic field. In the clean system the topology is influenced by the orientation of
the magnetic field and, in particular, the gapped or gapless nature of the states may lead
to different responses to disorder.  
As stated before, the difference of symmetry classes plus disorder gives rise to new universality classes.
Also, topology may be induced by quasiperiodicity, which leads to the expectation of new universality classes
(beyond the usual classification), as found in the one-dimensional case. In particular, one may expect 
interesting effects with the interplay of quasiperiodicity due to the presence of critical bulk states, 
and the existence of Majorana flat bands. The long-range nature of the quasiperiodic potential and the intrinsic
long-range nature of the gapless states may lead to an interesting competition. A distinction
between Anderson disorder (with moderate intensity) and quasidisorder is therefore interesting to consider, 
as shown in non-superconducting
systems, where for instance nodal points and nodal loops are affected differently by Anderson
disorder, or on a semimetal where imposing a quasiperiodic potential leads to flat bands.

\par The rest of the paper is organized as follows. Section \ref{sec:modelhamiltonian} introduces the model of the Hamiltonian and the topological properties of the clean system are discussed, first under a perpendicular and second under a parallel magnetic field, respectively in subsections \ref{sec:perpendicular} and \ref{sec:parallel}. In subsection \ref{sec:parallel} we derive the regions where the model is topological, and show that the topological regions are characterized by a Berry phase of $\pi$. In section \ref{sec:disorderedperpendicular} we present the results for the disordered model under a perpendicular magnetic field. We show that the introduction of Aubry-Andr\'e disorder leads to the appearance of topological phases in new regions. In \ref{sec:disorderedperparallel} we present the results for the disordered model under a parallel magnetic field. First we 
discuss the localization properties of the system in real space under different types of disorder, using the inverse participation ratio (IPR). We then turn to a mixed space description and discuss the evolution of the system as Anderson or Aubry-Andr\'e disorder are introduced. We show that the introduction of Aubry-Andr\'e disorder leads to the appearance of new regimes: for the $p$-wave superconductor, new gapless regimes with Majorana flat bands appear, 
and for the noncentrosymmetric superconductor, new regimes with unidirectional edge states appear.
We then obtain the Berry phase using twisted boundary conditions and show it is quantized to a value of $\pi$ for the quasidisorder induced flat bands. Identifying it as a topological invariant, we study two topological transitions and obtain the critical exponents $z$ and $\nu$, which we find to deviate from the known universality classes. Finally, using the IPR we study the nature of the eigenfunctions distinguishing between localized, single-fractal and multifractal regimes in the thermodynamic limit for both Anderson and Aubry-Andr\'e disorder. We conclude in section \ref{sec:conclusions}.
Three appendices discuss some further results on the disorder driven transitions under a perpendicular magnetic field in
Appendix \ref{detailsChern}, the influence of the dimensionality of the quasidisorder potential 
in Appendix \ref{pr2d} and the energy
spectra and density of states for the disordered noncentrosymmetric superconductor in Appendix \ref{sec:energyspectra}.

\section{\label{sec:modelhamiltonian}Model Hamiltonian}
In momentum space, the Bogoliubov-de Gennes (BdG) Hamiltonian matrix of the two dimensional model is written as
\begin{equation}
\mathcal{H}(\mathbf{k})=\left(\begin{array}{cc}
\xi(\mathbf{k})+\mathbf{B} \cdot \boldsymbol{\sigma} & \Delta(\mathbf{k}) \\
\Delta^{\dagger}(\mathbf{k}) & -\xi^{T}(-\mathbf{k})-\mathbf{B} \cdot \boldsymbol{\sigma}^{*}
\end{array}\right)
    \label{eqn:BdGHamiltonian}
\end{equation}
\noindent in a basis $(\boldsymbol{c}_{\mathbf{k} }^{\dagger},\boldsymbol{c}_{-\mathbf{k} }) = (c_{\mathbf{k} \uparrow}^{\dagger}, c_{\mathbf{k} \downarrow}^{\dagger}, c_{-\mathbf{k} \uparrow}, c_{-\mathbf{k} \downarrow})$ with $c_{\mathbf{k} \sigma}^{\dagger}$ ($c_{\mathbf{k} \sigma}$) the creation (annihilation) operator for an electron with momentum $\boldsymbol{k}=(k_x,k_y)$ and spin projection $\sigma$. In the BdG Hamiltonian, $\xi(\boldsymbol{k})=\epsilon_{\boldsymbol{k}} \sigma_{0}+\mathbf{s} \cdot \boldsymbol{\sigma}$, where  $\epsilon_{\mathbf{k}}=\left[-2t\left(\cos k_{x}+\cos k_{y}\right)-\mu\right] \sigma_{0}$ is the kinetic term, with $t$ the nearest-neighbour hopping integral and $\mu$ the chemical potential, $\mathbf{s}\cdot \boldsymbol{\sigma} = -\alpha(-\sin k_y, \sin k_x, 0)\cdot \boldsymbol{\sigma} = -\alpha\left[-\sin k_{y} \sigma_{x}+\sin k_{x} \sigma_{y}\right] $ is the Rashba spin-orbit term with $\mathbf{s}$ the spin-orbit vector. The term $\mathbf{B}\cdot \boldsymbol{\sigma}$ describes the Zeeman coupling of the electrons with an external magnetic field $\mathbf{B}$ and $\hat{\Delta}(\mathbf{k}) = \left[\Delta_{s}+\mathbf{d}(\mathbf{k}) \cdot \boldsymbol{\sigma}\right]\left(i \sigma_{y}\right)$ is the superconducting gap function. The pairing vector is chosen as $\mathbf{d} = d ( -\sin k_{y}, \sin k_{x} ,0)$, so that $d$ is the $p$-wave pairing amplitude and $\Delta_s$ is the $s$-wave pairing amplitude. The simultaneous existence of $s$ and $p$-wave terms is possible with a nonzero spin-orbit term, which breaks the parity symmetry.

\par The case of study is that of a system with periodic boundary conditions along the $x$ direction and open boundary conditions in the $y$ direction, such as in a cylinder geometry. Thus we can also write the Hamiltonian in a mixed space, $(k_x,y)$, where a Fourier transform to the reciprocal space is only done in the $x$ direction. In this case, for each value of $k_x$ the Hamiltonian matrix has a dimension $(4 \times N_y) \times (4 \times N_y)$, where $N_y$ is the number of sites in $y$. It is also of interest to write the Hamiltonian in real space. In this case the Hamiltonian matrix has dimension $(4 \times N) \times (4 \times N)$ with $N=N_x \times N_y$ the total number of sites and $N_x$, $N_y$ the number of sites in the $x$ and $y$ directions, respectively.
\par When $\mathbf{B}=0$, the system respects the time-reversal symmetry (TRS) $\mathcal{T}=(\sigma_0 \otimes \mathrm{i} \sigma_y)$ and the particle-hole symmetry (PHS) $\mathcal{P}=(\sigma_x \otimes \sigma_o)$ such that
\begin{equation}
\begin{split}
&\mathcal{P}\mathcal{H}(\mathbf{k})\mathcal{P^{\dagger}}=-\mathcal{H}^{*}(-\mathbf{k}), \\
&\mathcal{T}\mathcal{H}(\mathbf{k})\mathcal{T}^{\dagger}=\mathcal{H}^{*}(-\mathbf{k}),
\end{split}    
    \label{eqn:symmetries}
\end{equation}
and $\mathcal{T}^{2} = -1$, $\mathcal{P}^{2} = 1$. Therefore the Hamiltonian belongs to the DIII symmetry class, and if $|d| > |\Delta_s|$ the system  has a nontrivial $\mathbf{Z}_{2}$ number, displaying gapless counterpropagating Majorana edge states \cite{sato_fujimoto_2009, patricklee}.
\par For $\mathbf{B}\neq0$ the time-reversal symmetry is broken. The system exhibits different topological properties whether the applied magnetic field is perpendicular or parallel to the plane of the system, as will be now discussed.

\subsection{\label{sec:perpendicular}Perpendicular Magnetic Field}
\begin{figure}
	\centering
	%   \resizebox{0.8\textwidth}{!}{
	\includegraphics[width=0.92\linewidth]{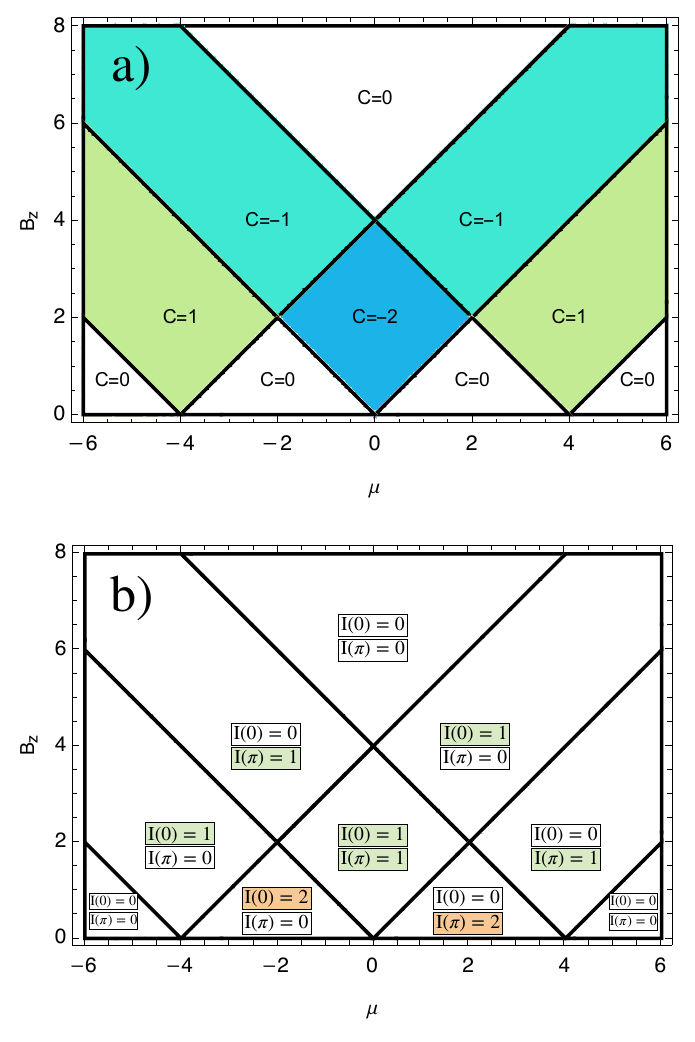}
	%    }
	\caption{Phase diagram for a) Chern number and b) winding number $I(k_y=0,\pi)$ as a function of $\mu$ and $B_z$, for $t=1$, $\Delta_s=0$, $d>0$.}
	\label{fig:Chern_Iky}
\end{figure}

Let us first consider the case in which the external magnetic field is perpendicular to the plane of the system,
$\mathbf{B} = (0,0,B_z)$. 
We have a gap closing point if one of the equations is satisfied \cite{sato_fujimoto_2009}:
\begin{equation}
\begin{split}
    (-4t-\mu)^2+\Delta^2_s=B^2_z, \\ \mu^2+\Delta^2_s=B^2_z, \\
    (4t-\mu)^2+\Delta^2_s=B^2_z.
    \label{eqn:gapclosingpoints}
\end{split}
\end{equation}
Eqs. \ref{eqn:gapclosingpoints} define the boundaries between regions in which the system has different topological properties. At the gap closing points the $D$ class system with broken time reversal symmetry undergoes topological transitions 
between gapped phases with different Chern numbers. The phase diagram of the system (indexed by the Chern number) is presented in Fig. \ref{fig:Chern_Iky}(a) for $t=1$,  $\Delta_s=0$, and $d>0$.

The regimes with a Chern number of zero and $B_z<2$, $0<|\mu|<4t$ exhibit edge states, besides having $  C=0$. This can be explained by one additional topological invariant. It can be defined noting that the Hamiltonian obeys a particle hole symmetry $\mathcal{P} = (\sigma_{x} \otimes \sigma_{0})$ with
\begin{equation}
\mathcal{P}\mathcal{H}(\mathbf{k})\mathcal{P}^{\dagger}=-\mathcal{H}^{*}(-\mathbf{k}).
    \label{eqn:phsymmetry}
\end{equation}
For the values $k_y=0$ and $k_y=\pi$, the Hamiltonian obeys $\mathcal{H}^{*}(-\mathbf{k})=\mathcal{H}(\mathbf{k})$ and thus anticommutes with $\mathcal{P}$, $\{\mathcal{H}(\mathbf{k}),\mathcal{P}\}=0$. Therefore the basis which diagonalizes $\mathcal{P}$ anti-diagonalizes the Hamiltonian.
A winding number $I(k_y)$ can then be defined as \cite{sato_fujimoto_2009}

\begin{comment}
For the values $k_y=0$ and $k_y=\pi$, the Hamiltonian obeys $\mathcal{H}^{*}(-\mathbf{k})=\mathcal{H}(\mathbf{k})$ and thus anticommutes with $\mathcal{P}$, $\{\mathcal{H}(\mathbf{k}),\mathcal{P}\}=0$. Therefore the basis which diagonalizes $\mathcal{P}$ anti-diagonalizes the Hamiltonian, and we may write
\begin{equation}
    \mathcal{H}(k_x)=\left(\begin{array}{cc}
0 & q(k_x) \\
q^{\dagger}(k_x) & 0
\end{array}\right)_{k_y=0,\pi}.
\end{equation}
A winding number $I(k_y)$ can then be defined as \cite{wen_zee_1989}
\end{comment}

\begin{equation}
\begin{split}
    I\left(k_{y}\right)=\frac{1}{4 \pi i} \int_{-\pi}^{\pi} d k_{x} \operatorname{tr}\big[q^{-1}(k_x) \partial_{k_{x}} q(k_x)- \\ q^{\dagger-1}(k_x) \partial_{k_{x}} q^{\dagger}(k_x)\big],\quad k_y=0,\pi,
\end{split}
\end{equation}
with
\begin{equation}
q(k_x)=\left(\begin{array}{cc}
-\epsilon_{\mathbf{k}}-B_{z}+i d \sin k_{x} & \Delta_{s}-i \alpha \sin k_{x} \\
-\Delta_{s}+i \alpha \sin k_{x} & -\epsilon_{\mathbf{k}}+B_{z}+i d \sin k_{x}
\end{array}\right)
\label{eqn:matrixq}
\end{equation}
the anti-diagonal block of the Hamiltonian matrix.
The values of $I(0)$ and $I(\pi)$ inside each phase are represented in Fig. \ref{fig:Chern_Iky}(b).
\par
\begin{figure}[h]
	\centering
	%   \resizebox{0.8\textwidth}{!}{
	\includegraphics[width=0.9\linewidth]{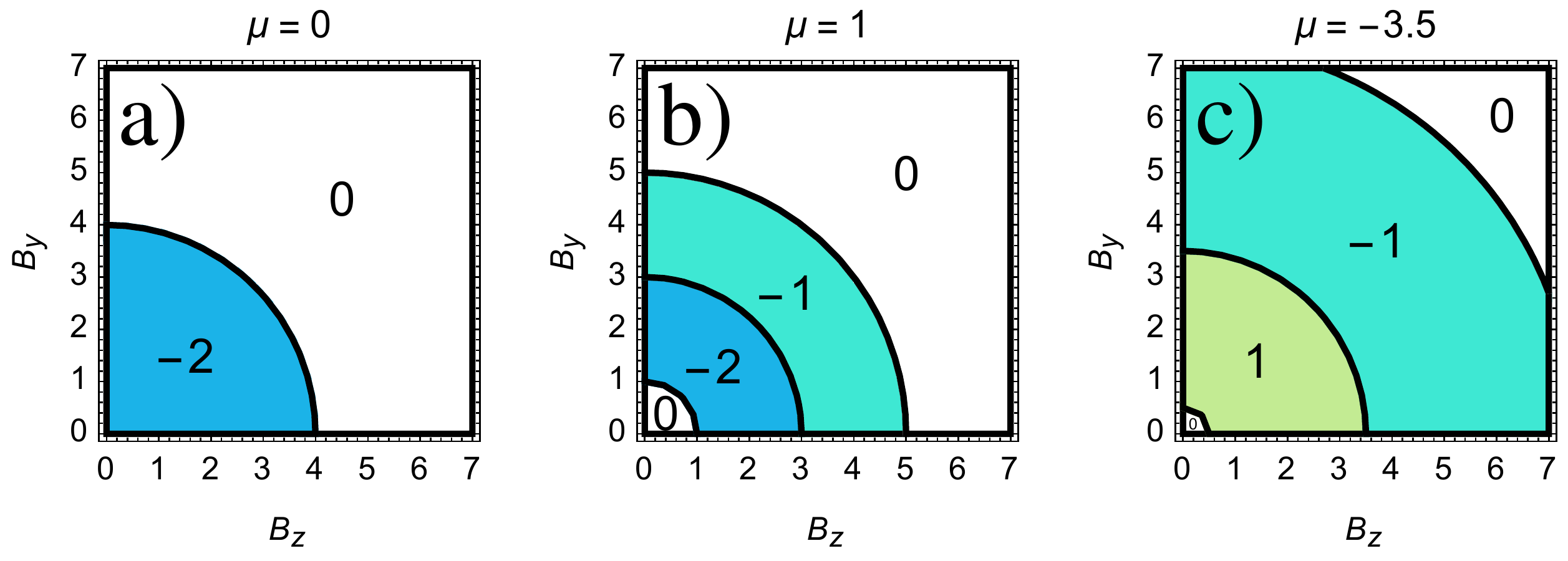}
	%    }
	\caption{Phase diagrams for ($B_z>0,B_y$), indexed by the Chern number, obtained numerically for a) $\mu=0$, b) $\mu=1$ and c) $\mu=-3.5$ for $\Delta_s=0$.}
	\label{fig:diagrambybz}
\end{figure}

\begin{figure}[h]
	\centering
	%   \resizebox{0.8\textwidth}{!}{
	\includegraphics[width=0.85\linewidth]{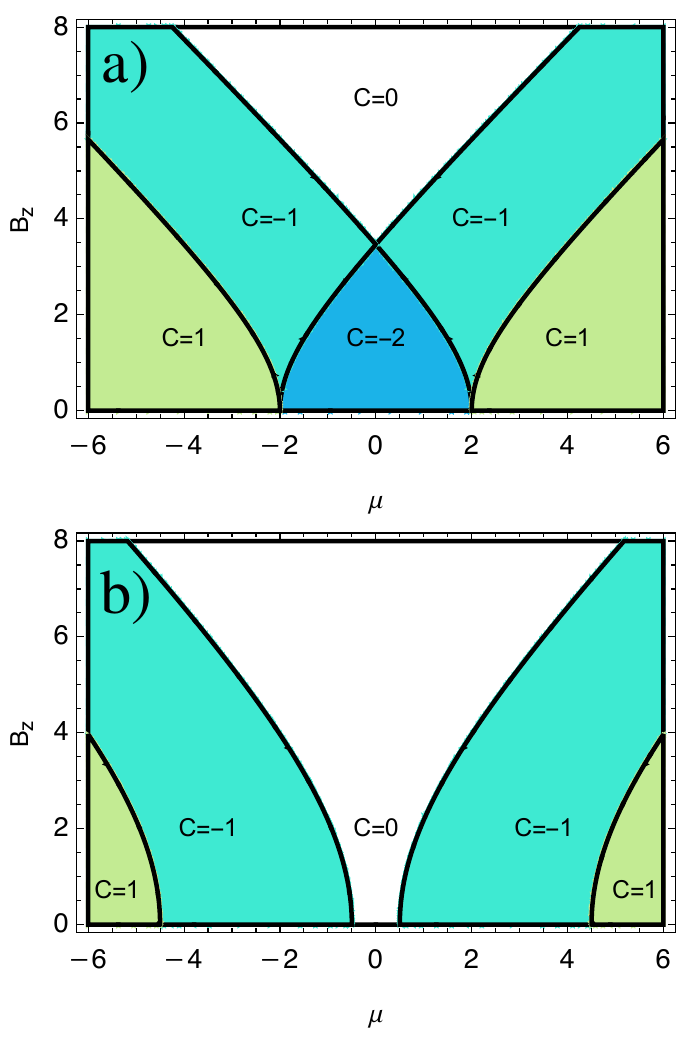}
	%    }
	\caption{
 Phase diagram indexed by the Chern number as a function of $\mu$ and $B_z$ (with $B_z>0$), for $t=1$, $\Delta_s=0$, $d>0$, and a) $B_y=2$ and b) $B_y=4.5$.}
	\label{fig:diagrambzby}
\end{figure}

 The invariant $I(k_y)$ loses its meaning if a finite magnetic field in the $y$ direction, $B_y$, is applied. However, we found that this is not true for the Chern number. Fig. \ref{fig:diagrambybz} shows phase diagrams indexed by the Chern number as a function of $B_z$ and $B_y$ for three different values of $\mu$. In this case the Chern number depends only on the value of $\sqrt{B^2_y+B^2_z}$. Also note that the diagrams only concern values of $B_z > 0$, excluding the points where $B_z=0$ and $B_y\neq 0$. In Fig. \ref{fig:diagrambzby} we present the phase diagram of the system as a function of $\mu$ and $B_z$ for constant values of $B_y$.

\subsection{\label{sec:parallel}Parallel Magnetic Field}
Now let us consider the case in which the applied magnetic field is parallel to the system,
$\mathbf{B} = (B_x,B_y,0)$. This could be realized, for instance, by threading a wire through the center of the superconductor in a cylindrical geometry.

Taking first the $s$-wave term $\Delta_s$ and the spin-orbit term $\alpha$ to be zero, the eigenvalues of the Hamiltonian are given by 
\begin{equation}
    E(\mathbf{k}) =  \pm \sqrt{z_1\pm 2 \sqrt{z_2}},
    \label{eqn:EigenvaluesPPB}
\end{equation}
with 
\begin{equation}
\begin{split}
z_1 = \mathbf{d}\cdot\mathbf{d} + \epsilon^2_{\mathbf{k}}+\mathbf{B}\cdot\mathbf{B}, \\
z_2 = \epsilon^2_{\mathbf{k}}(\mathbf{B}\cdot\mathbf{B})+(\mathbf{B}\cdot\mathbf{d})^2.
 \label{eqn:z1z2parallel}
\end{split}
\end{equation}

\par The gap closing points are solutions of the equation $z_1=2\sqrt{z_2}$, which is equivalent to the two equations being simultaneously satisfied:
\begin{equation}
    \begin{split}
    &\mathbf{d}\cdot\mathbf{d} + \epsilon^2_{\mathbf{k}}=\mathbf{B}\cdot\mathbf{B}, \\
    &(\mathbf{B}\cdot\mathbf{B})(\mathbf{d}\cdot\mathbf{d})=(\mathbf{B}\cdot\mathbf{d})^2.
    \end{split}
    \label{eqn:conditions}
\end{equation}
Eqs. \ref{eqn:conditions} simplify if we consider the magnetic field aligned with one of the axes. Let us then take the magnetic field aligned with the $y$ direction,  $\mathbf{B} = (0,B_y,0)$. In this case, the second equation simplifies to $\sin{k_y}=0$ which implies the bulk gap will close at $k_{y,0} = n\pi, n \in \mathbb{Z}$, provided there are values of $k_x$ that satisfy the equations

\begin{equation}
    d^2 \sin^2{k_x} + (-2t (\cos{k_x}\pm1)-\mu)^2 = B^2_y.
    \label{eqn:kxconditions}
\end{equation}

When the $p$-wave superconductor is in a gapless phase, and for a certain range of magnetic field, Majorana flat bands (MFBs) will appear in the system. This will be discussed next.
\par When finite spin-orbit $\alpha$ and $s$-wave pairing $\Delta_s$ terms are also considered, the flat bands will (for certain values of the magnetic field) acquire a slope, giving origin to unidirectional Majorana edge states (MESs). The appearance of such states is only possible with a gapless bulk, where a counter-propagating bulk current is created to cancel the edge current \cite{patricklee}.

\subsubsection{Flat bands: winding number and Berry phase quantization}
When the system is subject to an applied magnetic field, it no longer respects time-reversal symmetry. 
If the applied field has a generic form $\mathbf{B} = (B_x,B_y,0)$ we can, however, take $k_x$ as a fixed parameter of the Hamiltonian and find a set of symmetries that are only satisfied in the $y$ direction.
It is found that the Hamiltonian respects the symmetries:
\begin{equation}
\begin{split}
    \mathcal{T}_{k_y}^{-1}\mathcal{H}(k_x,k_y)\mathcal{T}_{k_y}= \mathcal{H}(k_x,-k_y), \\
\mathcal{P}_{k_y}^{-1}\mathcal{H}(k_x,k_y)\mathcal{P}_{k_y}= -\mathcal{H}(k_x,-k_y), \\
\end{split}
    \label{eqn:pseudosymmetries}
\end{equation}
where $\mathcal{T}_{k_y}=(\sigma_z \otimes \sigma_z)K$ and $\mathcal{P}_{k_y}=(\sigma_y \otimes \sigma_y)K$ are, respectively, defined as a "time-reversal-like" symmetry and a "particle-hole-like" symmetry \cite{patricklee} with $\mathcal{T}_{k_y}^2 = \mathcal{P}_{k_y}^2 = 1$ ($K$ is the complex conjugate operator). From these we can define a third chiral-like symmetry $\mathcal{S}_{k_y}=\mathcal{T}_{k_y}\mathcal{P}_{k_y}$:
\begin{equation}
\begin{split}
    \mathcal{S}_{k_y}^{-1}\mathcal{H}(k_x,k_y)\mathcal{S}_{k_y}= -\mathcal{H}(k_x,k_y). \\
\end{split}
    \label{eqn:pseudosymmetries2}
\end{equation}
\par
\begin{figure}[H]
	\centering
	\includegraphics[width=\linewidth]{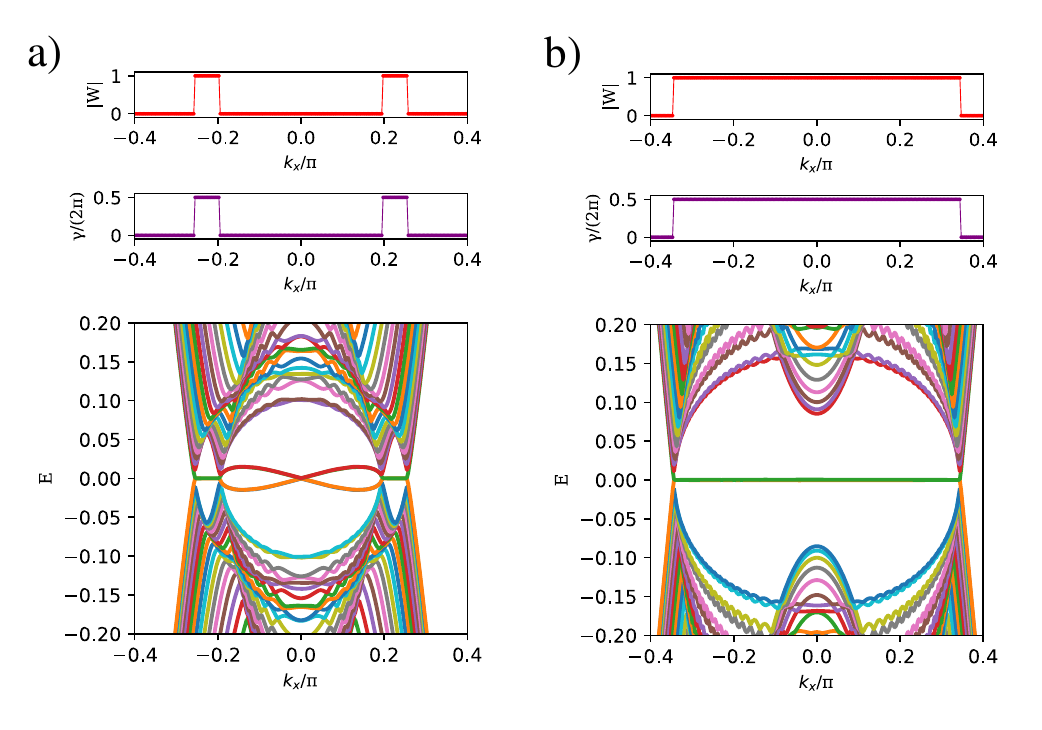}
	\caption{Energy spectrum, absolute value of the winding number $\mathcal{W}$ and Berry phase $\gamma$ normalized by $2\pi$, as a function of $k_x/\pi$. The values of the parameters are $t=1$, $d=1/6$, $\mu=-3.5$ and a) $B_y=d$, b) $B_y=3.5d$.}
	\label{fig:NBDIvsFlatbands}
\end{figure}

 Since we have that $\mathcal{T}_{k_y}^2 = \mathcal{P}_{k_y}^2 = 1$, the Hamiltonian belongs to the BDI symmetry class and, since the problem is effectively reduced to one dimension, the system can be characterized by an integer topological invariant. We can then write the Hamiltonian in the basis where $\mathcal{S}_{k_y}$ is diagonal, in which the Hamiltonian takes an anti-diagonal form. From here it is possible to obtain a winding number $\mathcal{W}$ at each value of $k_x$. It can be shown \cite{patricklee} that the winding number is calculated as
\begin{equation}
    \mathcal{W}(k_x) = \frac{i}{\pi} \left[ \log{\left(\frac{sgn(\mathcal{M}(k_y=0))}{sgn(\mathcal{M}(k_y=\pi))}\right)} \right]
\label{eqn:windingnumber}
\end{equation}
with
\begin{equation}
\begin{split}
 &\mathcal{M}\left(k_{x}, k_{y}\right) =  \\&\left[\mu+2t\left(\cos k_{x}+\cos k_{y}\right)\right]^{2}+d^{2} \sin ^{2} k_{x}-B_{y}^{2}+B_{x}^{2}.
    \label{eqn:mcaligrafico}
\end{split}
\end{equation}
In the regimes with $|\mathcal{W}|=1$ the system has a topological nature and Majorana flat bands appear, as is shown in 
Fig. \ref{fig:NBDIvsFlatbands}. These are protected by the chiral symmetry $\mathcal{S}_{k_y}$ as defined in Eq. \ref{eqn:pseudosymmetries2}.

The existence of topological flat bands may also be identified by a non-trivial Berry phase.
In general, the Berry phase can take any real value. In the presence of certain symmetry constraints, the Berry phase can become quantized to 0 or $\pi$ and carry topological information (at the value of $\pi$). This quantization can happen in the presence of inversion or chiral symmetries, also leading to the quantization of polarization \cite{benalcazar_bernevig_hughes_2017}. As the problem is reduced to one dimension, we can obtain a Berry phase $\gamma_B$ at each value of $k_x$, given by:
\begin{equation}
    \gamma_B(k_x)=i \int_{0}^{2 \pi}dk_y \langle\Psi(k_x,k_y) \mid \frac{\partial}{\partial k_y} \Psi(k_x,k_y)\rangle
    \label{eqn_BP}
\end{equation}
with $\Psi$ the ground-state wavefunction. The calculation is done numerically by discretizing the Brillouin zone \cite{benalcazar_bernevig_hughes_2017,xiao_berry_2010,resta_electrical_2010,fukui_chern_2005} in the $y$ direction. As is shown in 
Fig. \ref{fig:NBDIvsFlatbands}, we have found that in the regimes with $|\mathcal{W}|=1$, the Berry phase is also quantized to a value of $\pi$. 

\subsubsection{Domain of flat band existence: topological and gapless regions}
From Eq. \ref{eqn:windingnumber} it is found that $|\mathcal{W}|=1$ in the regimes where $\mathcal{M}\left(k_{x}, k_{y}=0\right)$ and $\mathcal{M}\left(k_{x}, k_{y}=\pi\right)$ have opposite signs. This is only possible if $|B_y|>|B_x|$, thus this is a necessary condition for the appearance of MFBs.
The flat band regions can be summarized in (with $\Tilde{B}^2 = B_{y}^{2}-B_{x}^{2}$):  
\begin{itemize}
    \item $(1)$ $\mu \geq 2t$
    \begin{equation}
\mathcal{D}_{+} > \Tilde{B}^2 > \mathcal{D}_{-}
    \label{eqn:region1}
    \end{equation}
\end{itemize}

\begin{itemize}
    \item $(2)$ $\mu \leq -2t$
        \begin{equation}
    \mathcal{D}_{-} > \Tilde{B}^2 > \mathcal{D}_{+}
     \label{eqn:region2}
    \end{equation}
    
    \item $(3)$ $-2t < \mu < 2t$
        \begin{equation}
    (\mathcal{D}_{+} > \Tilde{B}^2 > \mathcal{D}_{-})
     \vee  (\mathcal{D}_{-} > \Tilde{B}^2 > \mathcal{D}_{+})
    \label{eqn:region3}
    \end{equation}
\end{itemize}
where 
\begin{equation}
\mathcal{D}_{\pm}=\left[\mu+2t\left(\cos k_{x}\pm1\right)\right]^{2}+d^{2} \sin ^{2} k_{x}.
\label{eqn:Dpm}
\end{equation}

Eqs. \ref{eqn:region1}, \ref{eqn:region2} and \ref{eqn:region3} define the regions where the superconductor is in a nontrivial regime with $|\mathcal{W}|=1$, for a certain value of $k_x$. Furthermore, since MFBs can only appear in a gapless phase, the equations also define the regions where the bulk is gapless, as a function of the in-plane magnetic field.

\begin{figure}[H]
	\centering
	%   \resizebox{0.8\textwidth}{!}{
	\includegraphics[width=\linewidth]{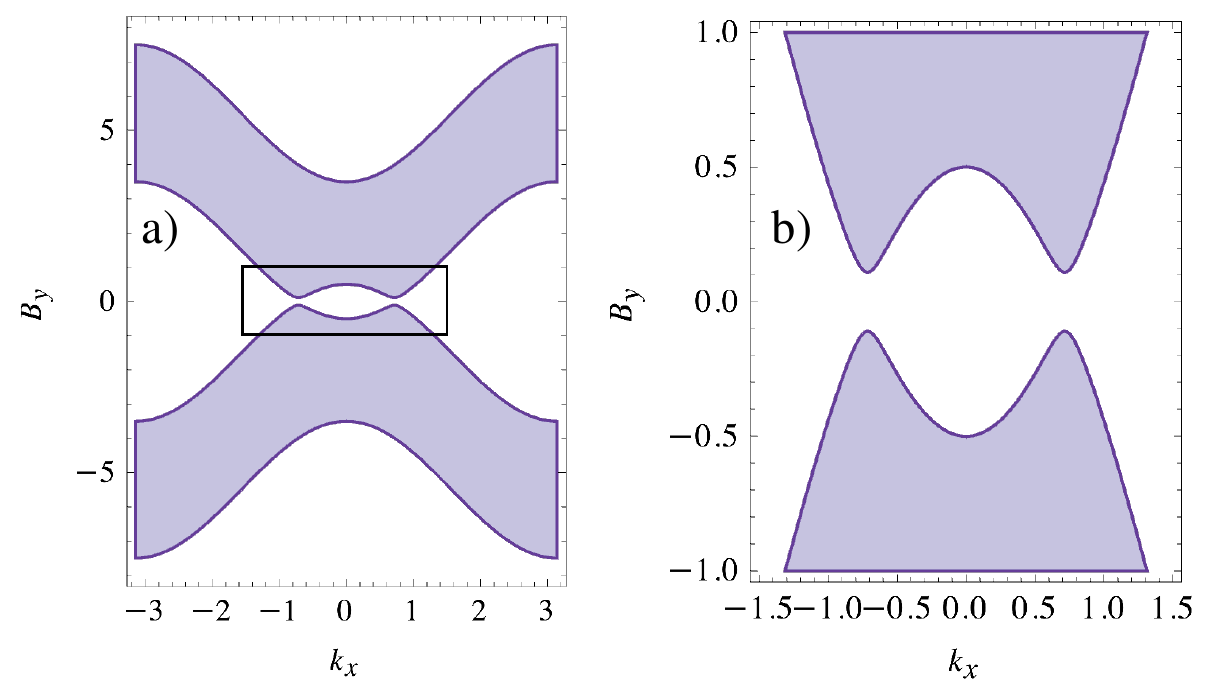}
	%    }
	\caption{a) Domain of existence of Majorana flat bands (shaded region) for $B_y$ vs. $k_x$ for the parameters  $t=1$, $d=1/6$, $\mu=-3.5$. b) Closeup of a) in the region $B_y \in$ [$-1,1$] and $k_x \in$ [$-1.5,1.5$].}
	\label{fig:fbregion}
\end{figure}

Note that the chiral-like symmetry that protects the flat bands is broken by either a non-zero $s$-wave pairing term $\Delta_s$ or a non-zero spin-orbit term $\alpha$. A finite perpendicular magnetic field $B_z$ is also found to break the chiral-like symmetry, leading to the absence of flat bands. 
If the flat band includes the point $k_x=0$, the addition of a finite $B_z$ will lead to the appearance of bands with a finite slope that cross at zero energy at $k_x=0$. Otherwise, the bands will be lifted to finite energy. 

\begin{figure*}
	\centering	\includegraphics[width=\linewidth]{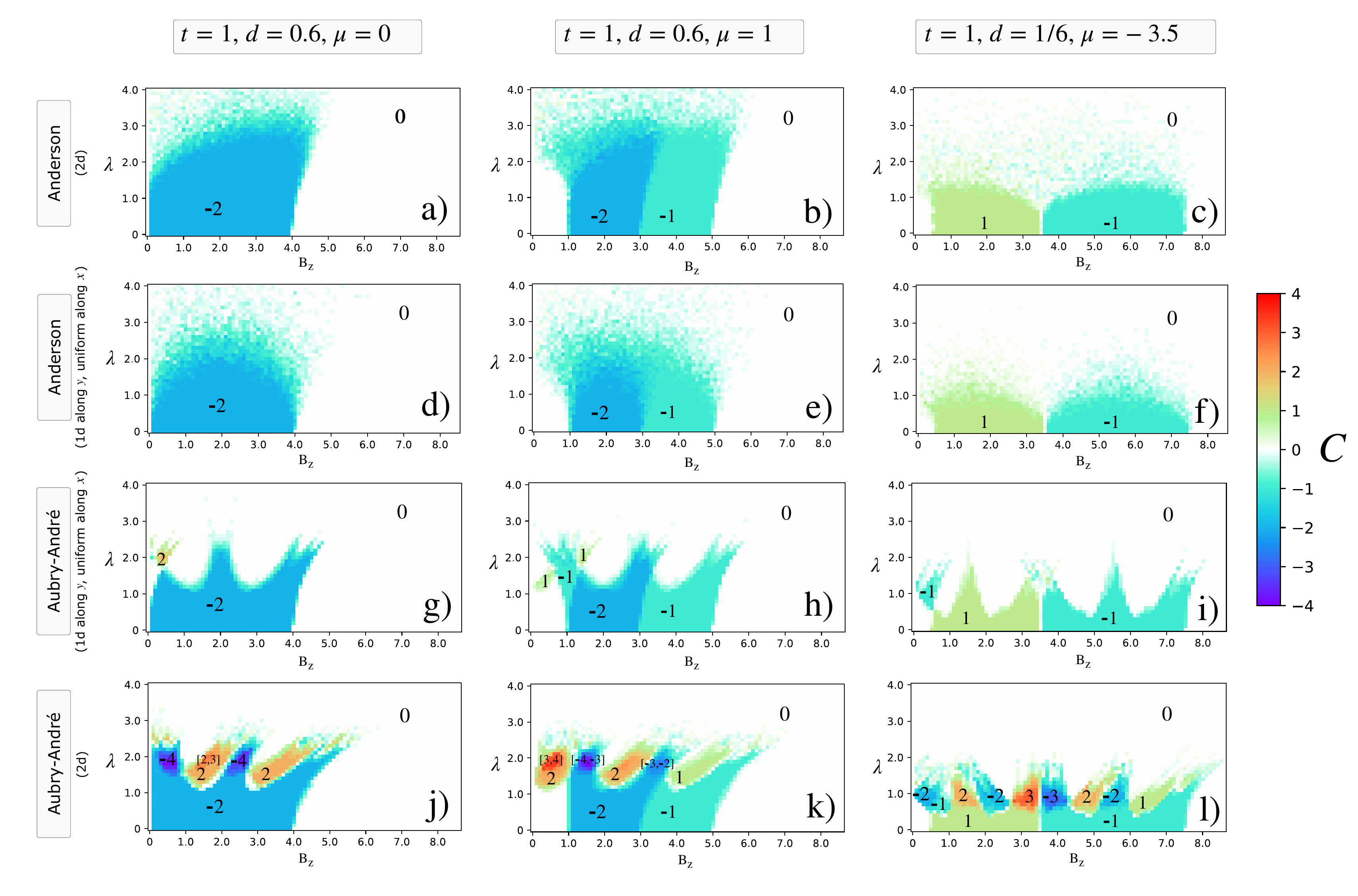}
	\caption{Phase diagrams indexed by the Chern number $C$ for a system with 20x20 sites, for several values of disorder strength $\lambda$ and perpendicular magnetic field $B_z$, obtained for an average over 10 disorder configurations. For Aubry-Andr\'e disorder, each random disorder configuration is obtained by selecting a random value of $\phi$. The first row with panels a)-c) concerns the case of Anderson disorder (2d), the second row with panels d)-f) concerns the case of Anderson disorder (1d along $y$, uniform along $x$), the third row with panels g)-i) concerns Aubry-Andr\'e disorder (1d along $y$, uniform along $x$), and the fourth row with panels j)-l) concerns Aubry-Andr\'e disorder (2d). The values of the parameters are $t=1$, $d=0.6$ and $\mu=0$ (left), $d=0.6$ and $\mu=1$ (middle), $d=1/6$ and $\mu=3d-4t=-3.5$ (right).}
\label{fig:ChernDiagrams}
\end{figure*}
\section{\label{sec:disorderedperpendicular}Disordered model under a perpendicular magnetic field}

We first want to investigate the effects of quasidisorder and disorder on the system subject to an applied magnetic field in the perpendicular direction, $\mathbf{B}=(0,0,B_z)$. 
Here we limit ourselves to the study of the system in real space and, 
to classify the topological nature of the system, the Chern number is obtained numerically \cite{chern_2013}.
We consider four different types of disorder potentials:
\begin{enumerate}
\item Anderson disorder (2d), where the disorder term is random at each site and varies with uniform probability within an interval:
\begin{equation}
\Lambda(x,y) \in [-\lambda,\lambda].
\label{eqn:AndersonDisorder_1}
\end{equation}

\item Anderson disorder (1d along $y$, uniform along $x$), where the potential is of the same type as described above but varies only along the $y$ direction, being uniform along the $x$ direction:
\begin{equation}
\Lambda(x,y) = \Lambda(y) \in [-\lambda,\lambda].
\label{eqn:AndersonDisorder}
\end{equation}

\item Aubry-Andr\'e disorder (1d along $y$, uniform along $x$), where the disorder term is a quasiperiodic potential of the form:
\begin{equation}
\Lambda(x,y) = \Lambda(y) = \lambda \cos(2 \pi \beta f(x,y) + \phi)
\label{eqn:AubryAndre}
\end{equation}
with $ f(x,y)$ a function of the lattice sites, $\beta=\frac{\sqrt{5}-1}{2}$ the inverse golden ratio, and $\phi$ a phase between $0$ and $2\pi$. Here we take $ f(x,y)=y$, so that the considered quasiperiodic potential is uniform in the $x$ direction.

\item Aubry-Andr\'e disorder (2d), where the disorder term is a sum of two quasiperiodic potentials of the form:
\begin{equation}
\Lambda(x,y) = \lambda \cos(2 \pi \beta x + \phi)+\lambda \cos(2 \pi \beta y + \phi)
\label{eqn:AubryAndre_xy}
\end{equation}
so that disorder potentials are introduced in both the $x$ and $y$ directions.
\end{enumerate}

\par In Fig. \ref{fig:ChernDiagrams} we show the phase diagrams indexed by the Chern number, for three different values of $\mu$ and $d$ (with $t=1$ in all cases) and for a system with size $20\times 20$. 
\par When Anderson disorder is introduced in the system (first row), the topological regimes are destroyed as the disorder strength is increased. There is, however, some difference in robustness as a function of the magnetic field. This is noticeable in Figs. \ref{fig:ChernDiagrams}(a) and (b), where we see that the robustness of the topological phases increases with the increase of $B_z$. In Fig. \ref{fig:ChernDiagrams}(a) and for a small region of magnetic field (for $B_z>4$) we observe reentrant topology as disorder is increased, 
as in Fig. \ref{fig:ChernDiagrams}(b), for lower values of magnetic field ($B_z<1$).
\par
The second row of the figure is obtained when 
disorder is considered with uniformity in the $x$ direction. Unexpectedly, the topological regions are to be less robust if compared with the previous case where Anderson disorder was considered with no modulation. Small traces of induced topology are observed for $B_z<1$ in Fig. \ref{fig:ChernDiagrams}(e) and $B_z<0.5$ in Fig. \ref{fig:ChernDiagrams}(f).
\par For Aubry-Andr\'e disorder uniform in the $x$ direction (third row) we obtain phase diagrams with well defined boundaries, and with induced topological regions. Here, the topological phases show an interesting and unexpected response to the increase of quasidisorder. There is a clear difference in robustness for different values of $B_z$, which originates the seemingly effect of “peaks” and “valleys” in the phase diagram, respectively at more robust and more vulnerable values of $B_z$. 
Induced topology is visible in panels g)-i), with topological transitions to finite values of $C$ happening at low and high values of the magnetic field with the increase of disorder.
\par The last row of Fig. \ref{fig:ChernDiagrams} concerns the case of two-dimensional Aubry-Andr\'e disorder. The introduction of disorder leads to the appearance of new topological regions, where several are characterized by values of $C$ that are not seen in the clean system, in the range of $[-4,4]$. Also, some regions appear where the Chern number oscillates within an interval between two integer values, without tending clearly to one of them.

We may argue that by adding disorder, local fluctuations of $\mu$ may lead to changes of the Chern number. This is particularly seen in the presence of quasidisorder. This suggests that the long-range quasiperiodicity resonates more with the calculation of the Chern number, that reflects the global structure of the states. 
Further details on the effect of disorder are shown in Appendix \ref{detailsChern}.

%\begin{figure}[H]
\begin{figure}
	\centering
	%   \resizebox{0.8\textwidth}{!}{
	\includegraphics[width=\linewidth]{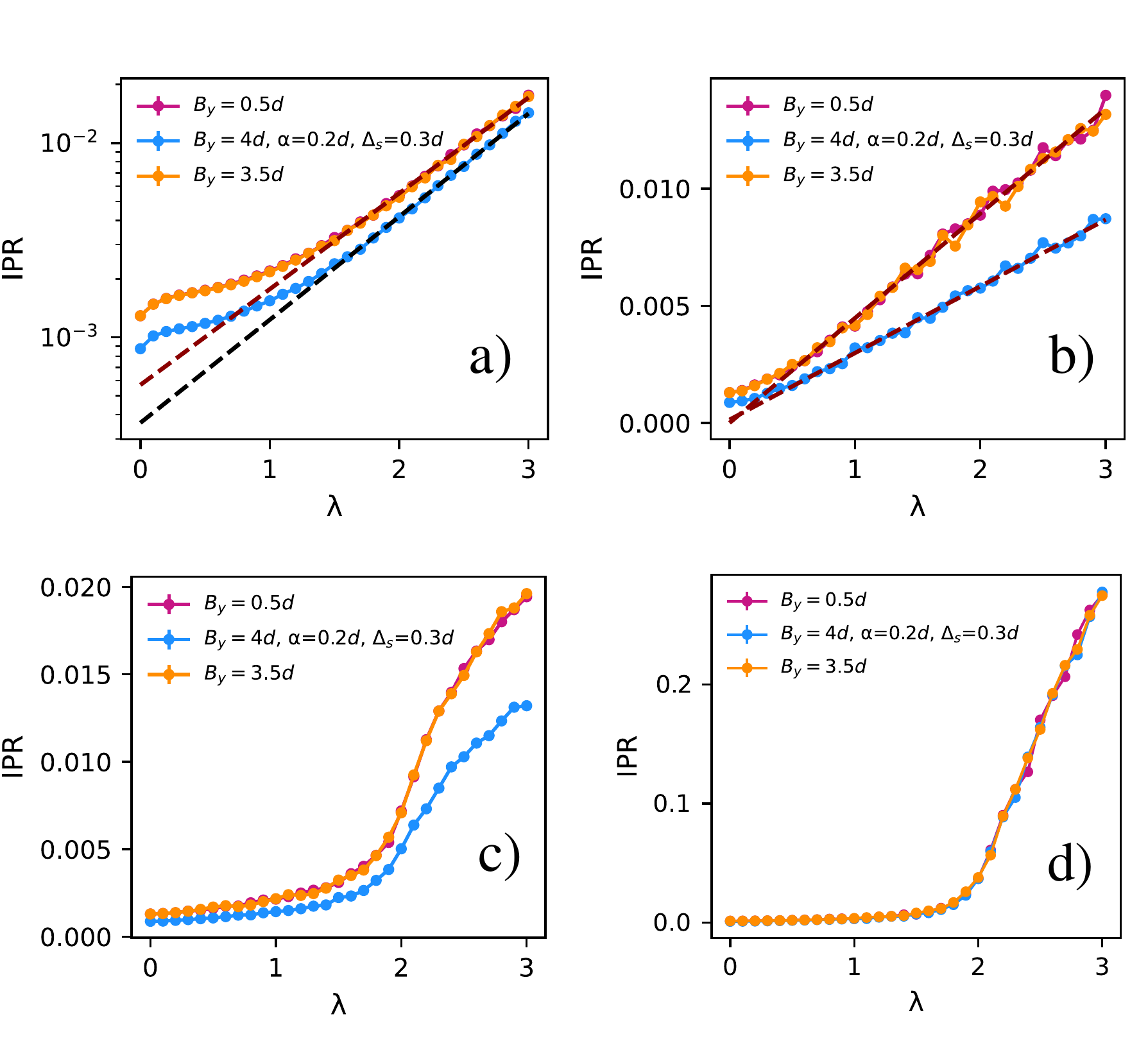}
	\caption{Average IPR of the whole system as a function of disorder strength, for a) Anderson disorder (2d), b) Anderson disorder (1d along $y$, uniform along $x$), c) Aubry-Andr\'e disorder (1d along $y$, uniform along $x$), d) Aubry-Andr\'e disorder (2d). The IPR is averaged over all eigenstates of the system in a given disorder configuration, and averaged over 10 disorder configurations. In a) fits are done to
	functions of the form $IPR=C_{1}\exp{C_{2}\lambda}$ in the range $\lambda \in [1.5,3]$, giving the values $(C_1,C_2)=(5.6\times10^{-4}, 1.13)$ for $B_y=0.5d$ and $B_y=3.5d$, and $(C_1,C_2)=(3.6\times10^{-4}, 1.22)$ for $B_y=4d$, $\alpha=0.2d$, $\Delta_s=0.5d$. In b) fits are done to functions of the form $IPR=C_{1}\lambda+C_2$ in the range $\lambda \in [0.5,3]$, giving the values $(C_1,C_2)=(4.4\times 10^{-3},5\times10^{-5})$ for $B_y=0.5d$ and $B_y=3.5d$, and $(C_1,C_2)=(2.8\times 10^{-3},1.4\times10^{-4})$ for $B_y=4d$, $\alpha=0.2d$, $\Delta_s=0.5d$.}
	\label{fig:IPRS}
\end{figure}

\section{\label{sec:disorderedperparallel}Disordered model under a parallel magnetic field}
We now introduce disorder on the system with an applied parallel magnetic field in the $y$ direction, $\mathbf{B}=(0,B_y,0)$. In section A we consider the system in real space, where the disorder term takes the same form as in the previous section (Eqs. \ref{eqn:AndersonDisorder_1}, \ref{eqn:AndersonDisorder}, \ref{eqn:AubryAndre},\ref{eqn:AubryAndre_xy}).
\par From section B onwards we study the system in a mixed $(k_x,y)$ space. In this case the disorder term is either of the form of 
Eq. \ref{eqn:AndersonDisorder} or Eq. \ref{eqn:AubryAndre},
with the potential varying in the $y$ direction and being the same for all $k_x$, as:

\begin{enumerate}
\item Anderson disorder:
\begin{equation}
\Lambda(y) \in [-\lambda,\lambda],
\label{eqn:AndersonDisordery}
\end{equation}

\item Aubry-Andr\'e disorder:
\begin{equation}
\Lambda(y) = \lambda \cos(2 \pi \beta y + \phi),
\label{eqn:AubryAndrey}
\end{equation}
where, as before, $\beta=\frac{\sqrt{5}-1}{2}$ is the inverse golden ratio and $\phi$ is a phase between $0$ and $2\pi$.
\end{enumerate}

\subsection{\label{sec:realspaceby}Localization properties in real space}
\par Here we briefly consider the system in real space and study its localization properties in three different regimes. 
We fix the parameter values as $t=1$, $d=1/6$, $\mu=-3.5$ and consider three different cases: the case of a $p$-wave superconductor for which a magnetic field $B_y=0.5d$ is added, such that the system is in a phase with a gapped bulk but gapless edge states; a $p$-wave superconductor with an added magnetic field of $B_y=3.5d$, where the system has a gapless bulk and is in the MFB regime; and a case of the noncentrosymmetric superconductor, with $B_y=4d$ and added $s$-wave pairing and spin-orbit terms, $\Delta_s=0.3d$ and $\alpha=0.2d$, where the system has a gapless bulk and unidirectional MESs.
\par To quantify the effects of disorder on the system's localization we use the inverse participation ratio, IPR. For a given eigenstate labeled by $m$, the IPR is defined as:
\begin{equation}
    \text{IPR}_{m}= \sum_{i}\left|\psi_{i}^{m}\right|^{4},
    \label{eqn:disorder_pr_ipr}
\end{equation}
with $\psi_{i}^{m}$ the wavefunction of the eigenstate $m$ at a site $i$. For perfectly localized states we have that $\text{IPR}_{m} \sim 1$ and for delocalized states $\text{IPR}_{m} \sim 1/N$. In Fig. \ref{fig:IPRS} we present results for the average IPR as a function of disorder for a system of size $N=N_x \times N_y=41\times 41$ \cite{sizes} and for the same types of disorder as before: 
Fig. \ref{fig:IPRS}(a) Anderson disorder (2d), Fig. \ref{fig:IPRS}(b) Anderson disorder (1d along $y$, uniform along $x$), 
Fig. \ref{fig:IPRS}(c) Aubry-Andr\'e disorder (1d along $y$, uniform along $x$), and Fig. \ref{fig:IPRS}(d) 
Aubry-Andr\'e disorder (2d). 

From observation of Figs. \ref{fig:IPRS}(a)-(d) we find four qualitatively different behaviours. In Fig. \ref{fig:IPRS}(a) 
(Anderson disorder) we see that the IPR shows an exponential-like behaviour for $\lambda>1.5$. A fit of the form $\text{IPR}=C_{1}\exp{C_{2}\lambda}$ is done in the range $\lambda \in [1.5, 3]$, and is presented in Fig. \ref{fig:IPRS}(a) in dashed lines. We find that for $\lambda>1.5$ the IPR follows an exponential behaviour closely, while for $\lambda<1.5$ there is a deviation from it. As disorder is increased, the low energy states become increasingly localized inside the bulk. From inspection of the wavefunctions we observed that the edge states quickly lose their structure for low values of disorder, although they do not become as quickly localized as the remaining bulk states. Accordingly, the IPR of these low energy states shows a slower increase than what is observed in Fig. \ref{fig:IPRS}(a).

\begin{figure}
	\centering
	%   \resizebox{0.8\textwidth}{!}{
	\includegraphics[width=\linewidth]{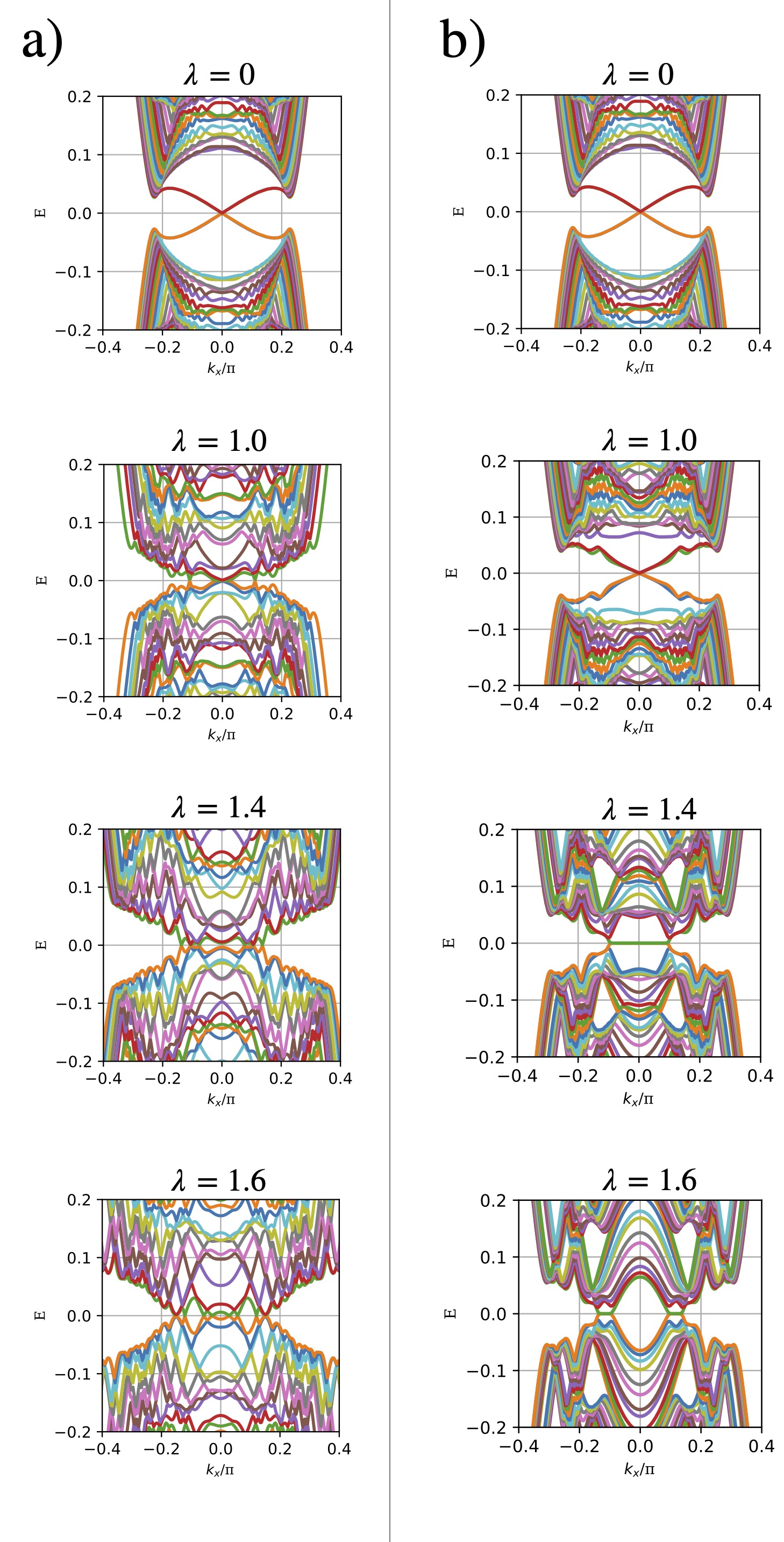}
	    	\caption{Energy spectra evolution with  a) Anderson disorder and b) Aubry-Andr\'e disorder, for $B_y = 0.5d$.}
	\label{fig:Evolution_By05d}
\end{figure}

\par For Anderson disorder along $y$ and with $x$ uniformity, we find a different localization behaviour. The IPR grows linearly with the increase of disorder, although with some fluctuations and a deviation for $\lambda < 0.5$. A fit of the form $\text{IPR}=C_{1}\lambda+C_2$ is done to the range $\lambda \in [0.5, 3]$ and presented in red dashed lines.
The change in behaviour in relation to \ref{fig:IPRS}(a) is a result of imposing periodicity in the $x$ direction on the disorder term, which unexpectedly causes the behaviour of the IPR to become linear.
The IPR of the low energy states follows a similar behaviour to what is seen for the average IPR.
By increasing disorder the low energy edge states are removed from the edges and localize inside the bulk, while remaining periodic in the $x$ direction.
\par In Fig. \ref{fig:IPRS}(c) (Aubry-Andr\'e disorder with $x$ uniformity) we see a threshold behaviour where a transition happens around $\lambda = 2$. For $\lambda < 2$ (approximately) there is a slow increase of the IPR, while for $\lambda > 2$ the IPR greatly increases. 
This resembles some known results: in the one-dimensional Aubry-Andr\'e model, where the system undergoes
an extended-localized transition at $\lambda=2t$, after which the average IPR shows a marked increase; for a one dimensional $p$-wave superconductor with an Aubry-Andr\'e potential this transition point changes to $\lambda$ = 2(t+d) with $d$ the $p$-wave pairing amplitude (when the chemical potential is taken as zero) \cite{cai_topological_2013,kitaevaa}.
For values of $\lambda$ before the transition, we observed that some bulk states acquire a critical like behaviour in the $y$ direction, while remaining periodic in the $x$ direction.
The low energy states are more robust to disorder if compared with the Anderson disorder cases, and are only removed from the system at the transition: after the threshold value of $\lambda$ there are no edge states in the system. 
\par Fig. \ref{fig:IPRS}(d) concerns the case of two-dimensional Aubry-Andr\'e disorder. 
The IPR shows again a threshold behaviour, and as in c) a transition is seen slightly below $\lambda=2$. 
However, the transition between two different regimes is abrupt in the IPR, and more closely resembles that of the one dimensional Aubry-Andr\'e chain. Also, unlike cases a)-c), the IPR follows the same behaviour for the three regimes considered. 
By a closer inspection of the IPR we see that this is only true for values of disorder over $\lambda=0.7$, as for $\lambda<0.7$ the noncentrosymmetric regime shows a consistently lower IPR, as before.
As disorder is increased for $\lambda>2$, states localize along both the $x$ and $y$ directions. 

The comparison between $1d$ Aubry-Andr\'e disorder and $2d$ Aubry-Andr\'e disorder, as well as a comparison between
a perpendicular and a parallel magnetic field, is detailed in Appendix \ref{pr2d}, with particular emphasis on the
existence of critical states and the apparent absence of a transition between extended and critical states, in contrast
to what is found in the one-dimensional case. 

\par Although results are not explicitly shown, the effect of edge disorder was also briefly studied, extending
previous results obtained for a time-reversal invariant system \cite{queiroz_schnyder_2014}. We considered both Anderson and Aubry-Andr\'e disorder potentials which were introduced locally at the edges at $y=0$ and $y=N_y$, varying along the $x$ direction (along the edge) also for a system of size $N=41\times 41$. We found that the bulk states and the system as a whole are almost unaffected by edge disorder, and the average IPR of the system remains nearly constant. However, the edge states are affected, and their behaviour depends on the type of disorder introduced. For Anderson disorder, the states localize continuously along the edges, while for Aubry-Andr\'e disorder there is also a threshold behaviour, similarly to what is presented in 
Fig. \ref{fig:IPRS}(c).

%\begin{figure}[H]
\begin{figure}
	\centering
	%   \resizebox{0.8\textwidth}{!}{
	\includegraphics[width=\linewidth]{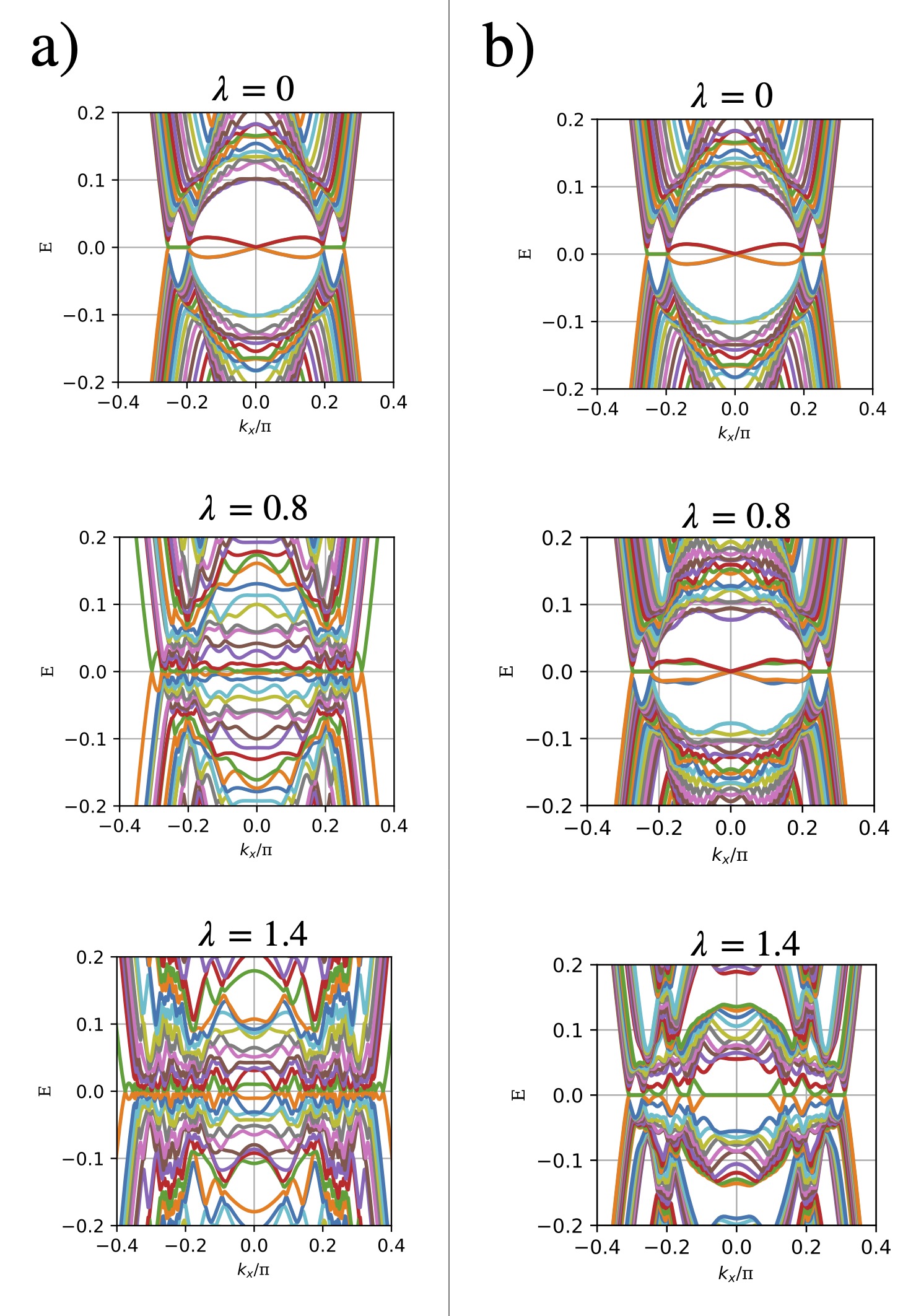}
	%    }
	\caption{
 Energy spectra evolution with  a) Anderson disorder and b) Aubry-Andr\'e disorder, for $B_y = d$.}
	\label{fig:Evolution_Byd}
\end{figure}

\subsection{\label{sec:disorderedperparallel_1}Energy spectra evolution and density of states}

\par We now consider the system in a mixed $(k_x,y)$ space, with finite width along $y$ and OBC. We fix the parameter values as $t=1$, $d=1/6$, $\mu=-3.5$, and $B_y=0.5d$ or $B_y=d$ (such that the system describes a $p$-wave superconductor) and obtain the evolution of the energy spectra for several values of disorder strength $\lambda$ for both Anderson and Aubry-Andr\'e disorder. Since the values of $t$, $d$ and $\mu$ will be kept constant we will now omit them. 

\par
\begin{figure}
	\centering
	%   \resizebox{0.8\textwidth}{!}{
	\includegraphics[width=\linewidth]{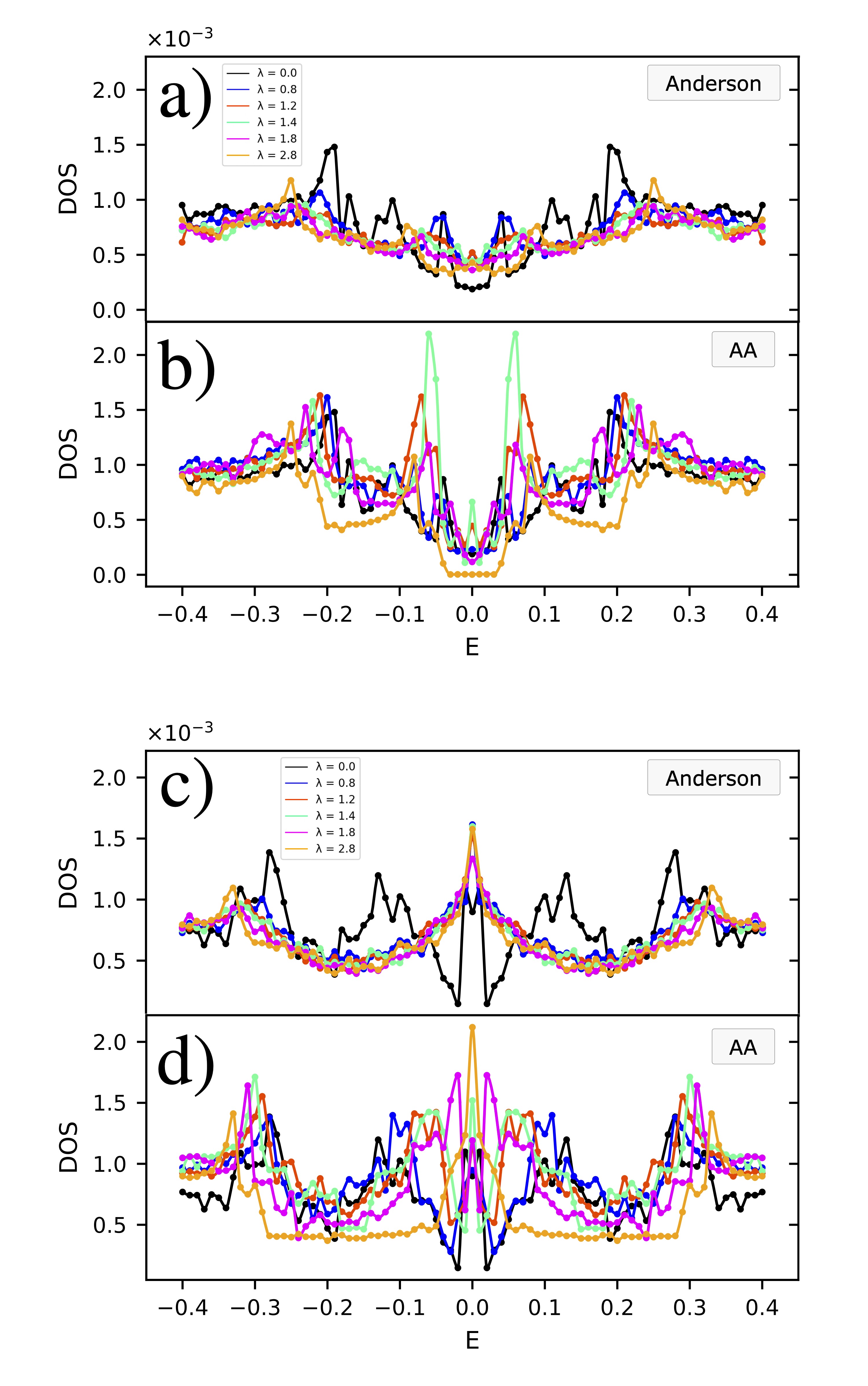}
	%    }
	\caption{
Density of states evolution with  a) Anderson disorder 
and b) Aubry-Andr\'e (AA) disorder, for $B_y=0.5d$ and with c) Anderson disorder 
and d) Aubry-Andr\'e disorder for $B_y=d$.}
	\label{fig:DOS_CS}
\end{figure}

 In Fig. \ref{fig:Evolution_By05d} we show the energy spectra for $B_y=0.5d$ with 
a) Anderson disorder and b) Aubry-Andr\'e disorder. The clean system has gapless edge states 
and the bulk gap is not closed by $B_y$. As Anderson disorder is increased, the edge states 
lose their structure and the bulk gap is closed. Accordingly, there is an increase in 
the density of states at $E=0$ and around zero energy as it can be seen in 
Fig. \ref{fig:DOS_CS}(a). 
\par Introducing quasidisorder, as seen in panel b), leads to a closing of the bulk gap with the 
appearance of new Majorana flat bands. As disorder is increased, the flat band then splits in 
two and disappears as a gap opens in the system for around $\lambda = 1.8$. 
The appearance of MFBs leads to an increase of the density of states at zero energy, as can be seen in 
Fig. \ref{fig:DOS_CS}(b) for the value of $\lambda=1.4$. At higher values of disorder, the system is gapped and the DOS at $E=0$ goes to zero. The reopening of the gap contrasts with what was found for Anderson disorder, where the bulk remains gapless as disorder is increased. We observed that the edge states inside the quasidisorder induced flat bands appear localized at both edges simultaneously. While the edge states of the clean system are localized symmetrically on both edges, the flat band states lose this symmetry and localize more near one of the edges if quasidisorder is present. Near the edge on which a given state appears less localized, there is also a deviation from the edge, and the state mostly localizes on the subsequent sites in $y$.

\par In Fig. \ref{fig:Evolution_Byd} the clean system with $B_y=d$ is in a gapless phase with both edge states and a range of $k_x$ supporting Majorana flat bands. As Anderson disorder is increased, the bulk remains gapless and there is a sharp increase in the density of states at zero energy, as the bulk states come from finite energies to lower energies. The sharp peak in the DOS observed at $E=0$ is reminiscent of the characteristic behaviour of a two-dimensional disordered superconductor
with broken time-reversal invariance 
in the thermal metal regime \cite{mildenberger_density_2007} in which the density of states displays a logarithmic divergence at zero energy. 

\par In Fig. \ref{fig:Evolution_Byd}(b) when Aubry-Andr\'e disorder is introduced, the edge states appear to be robust up until around $\lambda\approx0.8$. However, the MFBs which are present at $\lambda=0$ are more robust if compared with the edge states, with the band staying at zero energy but the initial range of $k_x$ hosting flat bands decreasing as $\lambda$ increases. Simultaneously, flat bands appear for new values of $k_x$, as is can be seen in the figure for $\lambda=1.4$, and accordingly, the density of states at zero energy increases. 
At higher values of disorder there is a collapse of states to lower energies and the density of states exhibits a peak at $E=0$ which is reminiscent of the behaviour found for Anderson disorder for the same parameter values. 
Contrary to what is observed in for a lower magnetic field, there is no opening of the bulk gap for 
larger values of $\lambda$. When quasiperiodic disorder is introduced, a gap will only open for larger 
values of $\lambda$ if the bulk was gapped prior to introducing disorder, as in 
Fig. \ref{fig:Evolution_By05d}, otherwise the bulk will remain gapless.

\par
\begin{figure}[H]
	\centering
	%   \resizebox{0.8\textwidth}{!}{
%	\includegraphics[width=0.9\linewidth]{MES.pdf}
	\includegraphics[width=0.9\linewidth]{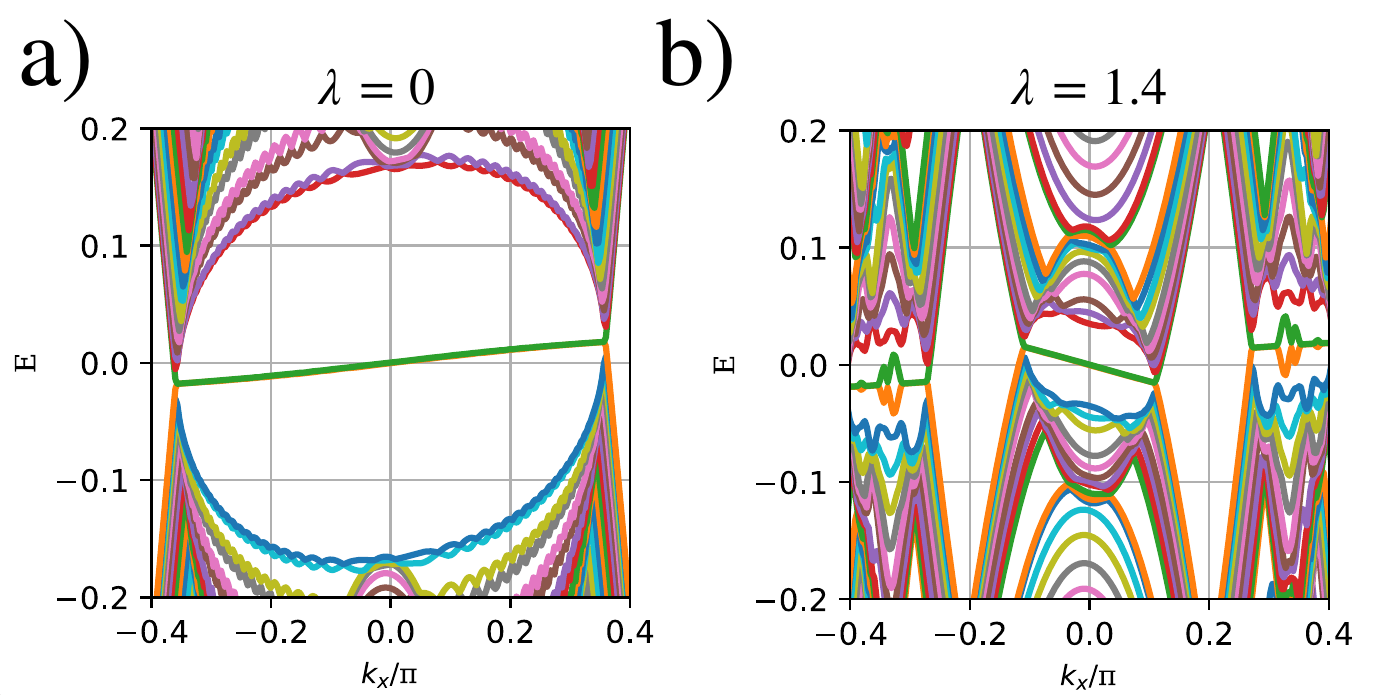}
	%    }
	\caption{Energy spectra evolution with Aubry-Andr\'e disorder for $B_y=4d$, $\alpha=0.2d$, $\Delta_s = 0.3d$ and a) $\lambda=0$, b) $\lambda=1.4$.}
	\label{fig:MES}
\end{figure}

 
Let us now consider the addition of finite values of $\alpha$ and $\Delta_s$. The addition of finite values of spin-orbit coupling and $s$-wave pairing potential breaks the chiral-like symmetry $\mathcal{S}_{k_y}$ (defined in Eq. \ref{eqn:pseudosymmetries2}) that protects the flat bands. As a result, the latter are lifted to a finite energy and the spectrum acquires a tilt.
For certain regimes of $B_y$, the noncentrosymmetric superconductor in the clean system shows unidirectional edge states. In such regimes, the addition of Aubry-Andr\'e disorder leads to the appearance of "flipped" unidirectional states in the system. This can be seen in Fig. \ref{fig:MES} for $\lambda=1.4$.

\subsection{\label{sec:disorderedperparallel_2}Topological nature of quasidisorder induced flat bands}
\par We want to investigate if the Majorana flat bands that arise in the presence of a quasiperiodic potential have a topological nature, such as is the case of the flat bands in the ordered system. Since the Berry phase was found to be quantized to a value of $\pi$ in the clean system in the region of flat bands, we calculate it here for the disordered case. The Berry phase $\gamma_B$ is obtained in real space using twisted boundary conditions. Considering a twisted boundary phase $\theta_y$ we have: 
\begin{equation}
    \gamma_B(k_x)=i \int_{0}^{2 \pi}d\theta_y \langle\Psi(k_x,\theta_y) \mid \frac{\partial}{\partial \theta_y} \Psi(k_x,\theta_y)\rangle
    \label{eqn_BPrealspace}
\end{equation}
where $\Psi$ denotes the ground-state many body wavefunction, which is given by the Slater determinant of the single particle wavefunctions. We can represent the ground state wavefunction by an $M \times N$ matrix $\mathbf{\Psi}^{\theta_y}$ where $N$ is the number of sites in $y$ and $M$ is the number of occupied states (negative energy states). Numerically, the twist variable is discretized into $L$ points between $0$ and $2\pi$, such that $\theta_y$ is constrained to take the values $\theta_{y,n} = \frac{2\pi}{L}n$, with $n$ an integer that goes from $0$ to $L-1$. A link variable can then be defined as $U(\theta_{y,n}) = \text{det}\left[\mathbf{\Psi^{\dagger}}_{\theta_{y,n}}\mathbf{\Psi}_{\theta_{y,n+1}}\right]$, and the Berry phase is obtained as
\begin{equation}
\gamma_B = -\mathrm{i} \sum_{n=1}^{L}\log{U(\theta_{y,n})}.
\label{eqn:BPrealspace}
\end{equation}

\begin{figure}[H]
	\centering
	%   \resizebox{0.8\textwidth}{!}{
	\includegraphics[width=\linewidth]{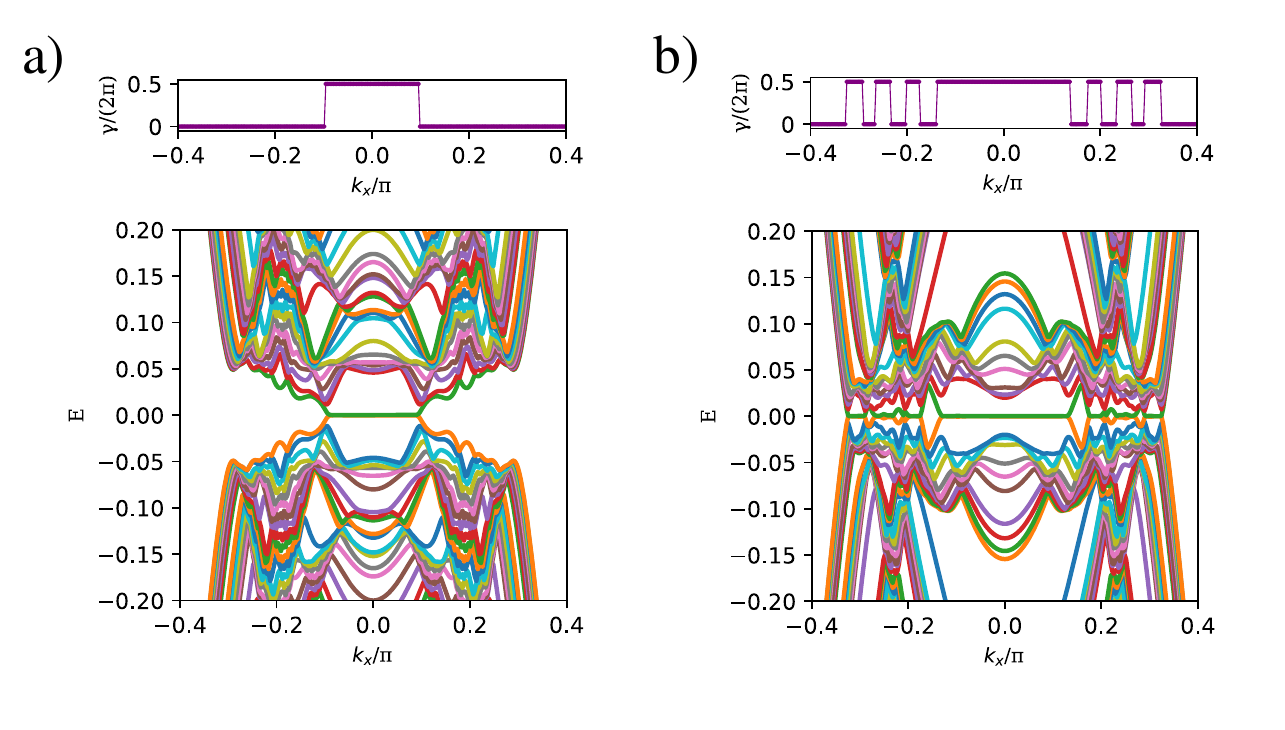}
	%    }
	\caption{Energy spectrum and Berry phase $\gamma$ normalized by $2\pi$, as a function of $k_x$. The values of the parameters are $t=1$, $d=t/6$, $\mu=3d-4t$ and a) $B_y=0.5d$, $\lambda=1.4$ b) $B_y=d$, $\lambda=1.6$, with $\lambda$ the strength of the quasiperiodic Aubry-Andr\'e potential.}
	\label{fig:MFBs}
\end{figure}

\par We find that at the values of $k_x$ where Majorana flat bands appear the Berry phase is quantized to $\pi$, as shown in 
Fig. \ref{fig:MFBs}. We have also found that, for the considered system sizes, the values of $k_x$ at which the Berry phase is quantized to $\pi$ are independent of the phase $\phi$ in the Aubry-Andr\'e potential.
\par
\begin{figure}[H]
	\centering
	%   \resizebox{0.8\textwidth}{!}{
	\includegraphics[width=\linewidth]{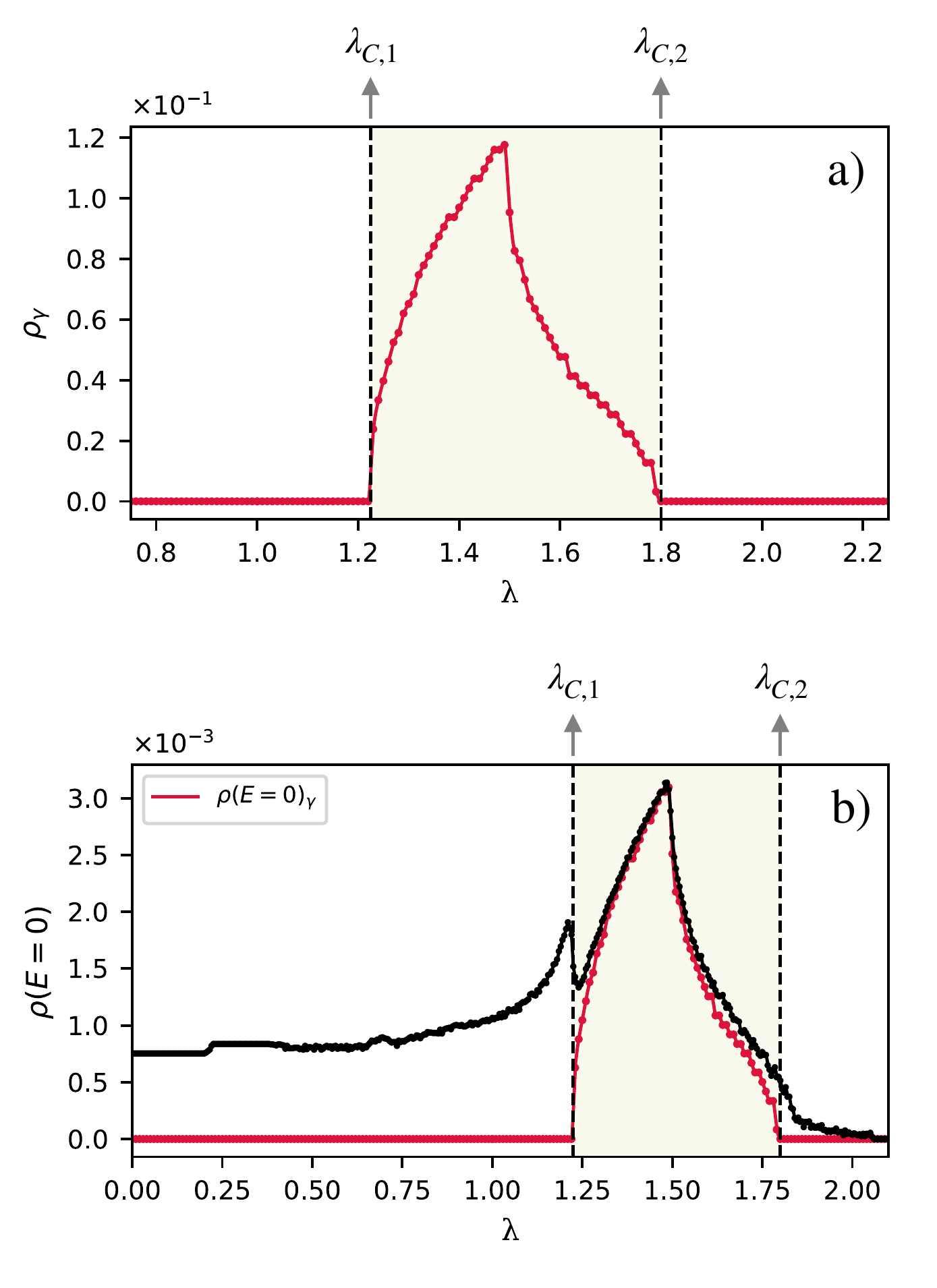}
	%    }
	\caption{a) Values of $\rho_{\gamma}$ for the case $t=1$, $d=t/6$, $\mu=-3.5$ and $B_y=0.5d$ vs. quasidisorder strength $\lambda$. Obtained for a system with $76$ sites in $y$. b) Value of the DOS at $E=0$ for the same parameter values as in a), vs. quasidisorder strength $\lambda$, and the contribution for $\rho(E=0)$ which comes from the Majorana flat bands in the corresponding regime.}
	\label{fig:GammaDensityBy05d}
\end{figure}

 To quantify the induced bands at zero energy and study the transition to a $\pi$-quantized Berry phase, we use the concept of Majorana pair density, defined as \cite{sedlmayr_aguiar_hualde_bena_2015}:
\begin{equation}
    \rho_{\gamma} = \frac{N_{\gamma}}{N_{k}},
    \label{eqn:MPDensity}
\end{equation} 
where $N_{k}$ is the number of discrete points of $k_x$ taken inside the interval $\left[-\pi,\pi\right]$, and 
$N_{\gamma}$ is the number of such points which support MFBs at the edges. Numerically it is more convenient to 
consider the number of $k_x$ points for which the Berry phase is quantized to $\pi$, $N_{\pi}$, since it was found 
that $N_{\pi}=N_{\gamma}$. A transition from $\rho_{\gamma}=0$ to $\rho_{\gamma}\neq0$ then signals a transition 
from a trivial to a topological regime ($\pi$-quantized Berry phase). Fig. \ref{fig:GammaDensityBy05d}(a) shows 
the evolution of $\rho_{\gamma}$ as a function of quas-idisorder strength for the case $t=1$, $d=t/6$, $\mu=-3.5$ and $B_y=0.5d$, and for the range $\lambda \in [1,2]$. A transition $\rho_{\gamma}=0 \rightarrow \rho_{\gamma}\neq0$
 occurs between $\lambda=1.22$ and $\lambda=1.23$ at a certain critical value $\lambda_{C,1}$. 
The value of $\rho_{\gamma}$ grows until $1.49 \pm 0.01$ when the flat band splits in two and the behaviour of $\rho_{\gamma}$ changes, with an abrupt change in the sign of the second derivative. A second transition occurs between $1.79$ and $1.8$, at a critical value $\lambda_{C,2}$, where $\rho_{\gamma}$ becomes zero.

\par In Fig. \ref{fig:GammaDensityBy05d}(b) the density of states at zero energy $\rho(E=0)$ (normalized by the system size) is shown, for the same parameters as in Fig. \ref{fig:GammaDensityBy05d}(a) and for $N_y=76$, along with the corresponding contribution for the zero energy density of states which comes from the MFB, $\rho(E=0)_{\gamma}$. Inside the topological phase, which is highlighted, we can see that the finite value of $\rho(E=0)$ observed in the system with OBC comes almost entirely from the presence of flat bands.
\begin{table}[H]
\centering
\begin{tabular}{ccc}
\hline 
$N_y$ & $\lambda_{C,1}$   & $\lambda_{C,2}$ \\ \hline
76 & $1.225 \pm 0.005$ & $1.800 \pm 0.005$ \\
100 & $1.215 \pm 0.005$ & $1.775 \pm 0.005$ \\
175 & $1.230 \pm 0.005$ & $1.805 \pm 0.005$ \\
200 & $1.220 \pm 0.005$ & $1.800 \pm 0.005$ \\
400 & $1.230 \pm 0.005$ & $1.805 \pm 0.005$ \\
800 & $1.225 \pm 0.005$ & $1.805 \pm 0.005$ \\
\hline 
\end{tabular}
\caption{Values of the critical points $\lambda_{C,1}$ and $\lambda_{C,2}$, for the parameter values $t=1$, $d=t/6$, $\mu=-3.5$ and $B_y=0.5d$ and for the system sizes $\{76,100,175,200,400,800\}$.}
\label{tab:lambdacritico}
\end{table}

Table \ref{tab:lambdacritico} shows the values of $\lambda_{C,1}$ and $\lambda_{C,2}$ for several system sizes, obtained for random values of the phase $\phi$ in the Aubry-Andr\'e potential, where the uncertainty is taken as the minimum interval considered between values of $\lambda$. Is is found that the values of the critical points show little variation with the system size, and we also found that the critical points are independent of $\phi$ for the system sizes considered.

\subsection{\label{sec:disorderedperparallel_3}Scaling of the density of states: critical exponents}
\subsubsection{A detour to the clean system}
Let us first briefly consider the clean system, without disorder. For the clean case, it is possible to obtain the values of the dynamical exponent $z$ and of the critical exponent $\nu$ analytically, for the transition that occurs as $B_y$ is increased, corresponding to a transition from a winding number of 0 to 1 or a Berry phase of 0 to $\pi$.
Here we consider the case of $\mu<-2t$ (such that the topological phase is within the region described by Eq. \ref{eqn:region2}).
At the topological transition to a gapless phase, the gap closing points in $k_x$, $k_{x,0}$, are given by
\begin{equation}
    k_{x,0} = \pm \arccos{\left[-\frac{2(t \mu +2 t^2)}{-d^2+4t^2} \right]} + 2n\pi, n \in \mathbb{Z}.
    \label{eqn:gapclosingpointss}
\end{equation}
The values of $k_y$ for which the gap closes are given by $k_{y,0}=n\pi$, $n \in \mathbb{Z}$ (general solution). In this case the transition happens at $k_{y,0}=2n\pi$, $n \in \mathbb{Z}$. The gap closes at a critical value of the magnetic field, $B_{y_C}$, which, fixing $k_{y}=k_{y,0}$, is defined from the value of $k_{x,0}$ as
\begin{equation}
    B^2_{y_C} = \left[\mu+2t\left(\cos k_{x,0}+1\right)\right]^{2}+d^{2} \sin^{2} k_{x,0}.
    \label{eqn:CriticalBy}
\end{equation}
We can now first expand the expressions for the bulk energy around $k_{x,0}$ to find the dependence of the energy on $k_x$. We only need to consider the first positive energy band, $E_{+}(k_x)$. Taking $k_y=k_{y,0}$ and expanding around $k_x = k_{x,0}$ we find
\begin{equation}
\begin{split}
    E_{+}(k_x) \propto (k_x - k_{x,0}),
\end{split}
    \label{eqn:Energies_kx0}
\end{equation}
implying a value of the dynamical exponent $z=1$ for the transition.
We can now take $k_x=k_{x,0}$ and see how the gap closes as a function of $B_y$. We find
\begin{equation}
\begin{split}
    E_{+}(k_x=k_{x,0}) = \left| |B_{y_C}| - |B_{y}| \right|.
\end{split}
    \label{eqn:Energies_By}
\end{equation}
Near a quantum phase transition as a critical point $\lambda_C$ is approached, the gap behaves as $\Delta \sim |\lambda-\lambda_C|^{z\nu}$, therefore at $k_x=k_{x,0}$ the gap vanishes linearly, with an exponent $z\nu = 1$. Since $z=1$, this implies $\nu=1$, and
\begin{equation}
z=1, \quad \nu=1.
    \label{eqn:z_nu}
\end{equation}

\subsubsection{Quasidisorder: numerical calculation of the critical exponents}
Around a critical point, the density of states $\rho(E)$ follows \cite{kobayashi_ohtsuki_imura_herbut_2014}
\begin{equation}
    \rho(E)=\delta^{(D-z)\nu}f(|E|\delta^{-z\nu}),
    \label{eqn:DOSscaling7}
\end{equation}
with $D$ the dimension of the system (here $D=2$), $\delta = \frac{|\lambda-\lambda_C|}{\lambda_C}$ the normalized distance to the critical point $\lambda_C$, and $f$ a scaling function. Right at the critical point, when $\delta=0$, the DOS behaves as
\begin{equation}
    \rho(E)\sim|E|^{\frac{D}{z}-1}.
    \label{eqn:DOSscaling8}
\end{equation}
From the behaviour of the density of states near the phase transition and using Eqs. \ref{eqn:DOSscaling7} and
\ref{eqn:DOSscaling8} it is possible to obtain the values of the critical exponents numerically.
\par Here we study a system with $N_y=800$ sites in $y$ and consider the obtained critical values $\lambda_{C,1}=1.225$ and $\lambda_{C,2}=1.805$ (as shown in table \ref{tab:lambdacritico} for this system size). A fit of the form of Eq. \ref{eqn:DOSscaling8} for the density of states at the critical points, done in the interval E $\in [0.005,0.025]$, gives the values of the critical exponents $z=1.27\pm0.04$ for the first transition and $z=1.23\pm0.03$ for the second transition. To determine the value of $\nu$ we take values of $\lambda$ inside the topological (gapless) phase, $\lambda > 1.225$ and $\lambda < 1.805$, and obtain the density of states close to zero energy. For small values of $\delta$ and close to zero energy a collapse of the scaled values of the density of states according to 
Eq. \ref{eqn:DOSscaling7} is expected.

%\begin{figure}[H]
\begin{figure}
	\centering
	%   \resizebox{0.8\textwidth}{!}{
	\includegraphics[width=\linewidth]{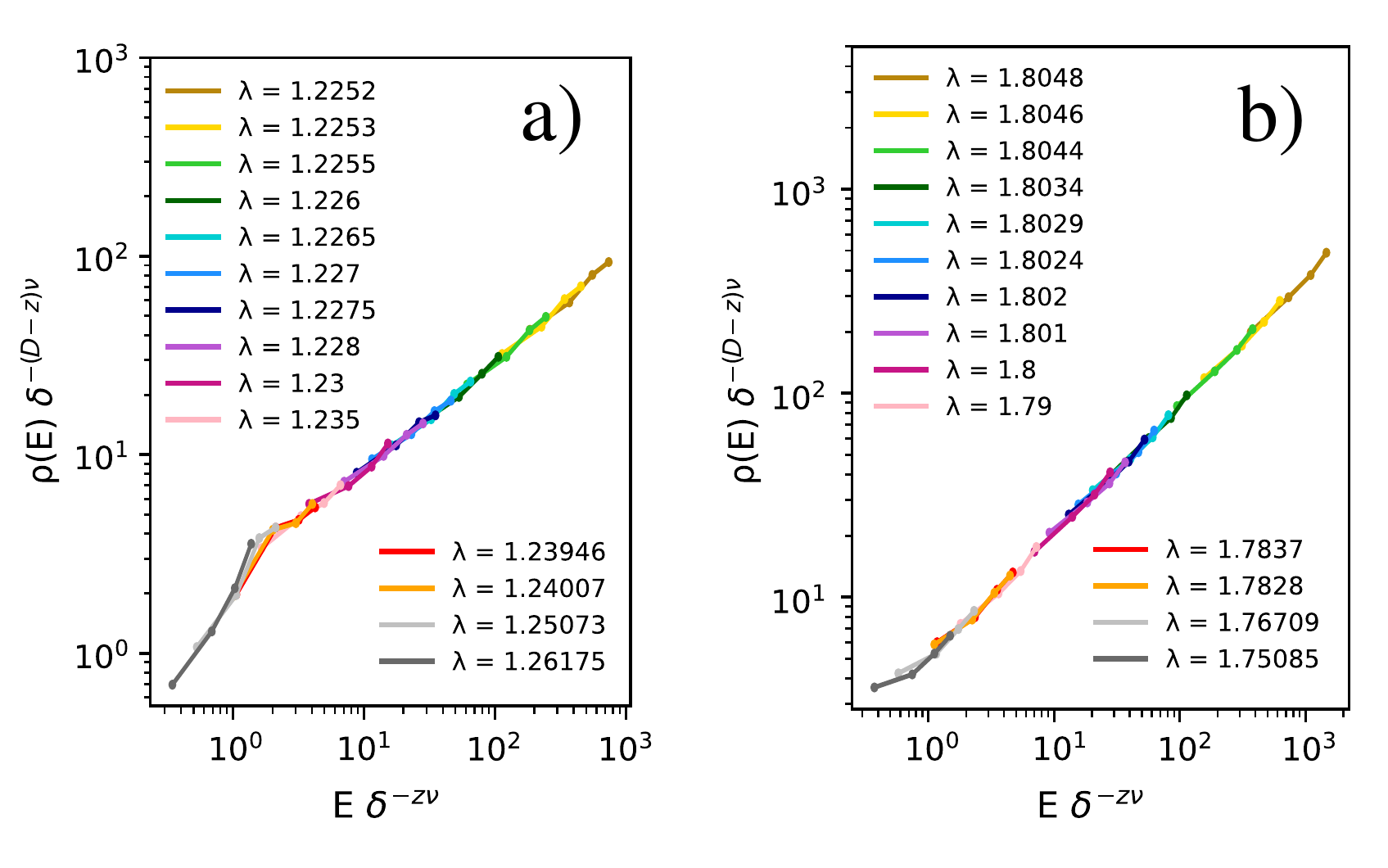}
	%    }
	\caption{Density of states for $E \in [0.005,0.025]$ and several values of $\lambda$ close to the critical values, for a) $\lambda_{C,1}=1.225$ and b) $\lambda_{C,2}=1.805$, scaled according to Eq. \ref{eqn:DOSscaling7} for a) $z=1.27$ and $\nu=0.95$ and b) $z=1.23$ and $\nu=1.00$.}
	\label{fig:DOS-Saling-Nu_Disorder}
\end{figure}

\par In Fig. \ref{fig:DOS-Saling-Nu_Disorder} we show the results for the scaled density of states for: a) values close to the first transition at $\lambda_{C,1}=1.225$, and b) values close to the second transition at $\lambda_{C,2}=1.805$. The density of states shows a collapse for a) $z=1.27$ and $\nu=0.95$ and b) $z=1.23$ and $\nu=1.00$. 
\par The quantum phase transitions in the disordered regime are therefore in a different universality class than that of the clean case, which was found to behave with $z=\nu=1$. The obtained values also differ significantly from the known results for the Anderson or the Aubry-Andr\'e transitions in one dimension, the first belonging to an universality class with with $\nu=2$ and $z=2/3$, and the second case with critical exponents 
$\nu=1$ and $z=2.375$ \cite{cestari_critical_2011}. Recent results show that for a one dimensional system with $p$-wave superconductivity subject to an Aubry-Andr\'e potential the quasidisorder driven transitions also deviate from the normal Aubry-Andr\'e class. For the localized-critical transition line and when the $p$-wave pairing term is finite, the correlation length exponent has been obtained as $\nu = 0.997$ and the dynamical exponent as $z = 1.373$ in \cite{tong_meng_jiang_lee_neto_xianlong_2021}, and as $\nu= 1.000$, $z = 1.388$ in \cite{lv_quantum_2022}. Note, however, that the referred results are for $D=1$ while we are studying a two dimensional system, and concern systems with no applied magnetic field. Nevertheless, one could say that the aforementioned results make it so that deviations from the known universality classes are also expected for transitions in the system at study.
Up to numerical errors, the values of $\nu$ obtained for the disordered driven transitions coincide with that of the Aubry-Andr\'e transition; nevertheless the value of $z$ deviates from that of the known classes, which suggests these transitions belong to new universality classes.
The identified transitions, where Marojana flat bands appear as a result of a quasidisorder induced gap closing, and the subsequent opening of the bulk gap, are found to happen for other values of the imposed parameters. Considering the values of the parameters $\mu$, $t$ and $d$ are such that the topological regions of the superconductor are described by Eq. \ref{eqn:region2}, then as long as $B_y<B_{y,C}$ (when the bulk is gapless) with $B_{y,C}$ defined as in Eq. \ref{eqn:CriticalBy}) the same type of transitions will take place with the increase of $\lambda$. 

\begin{table}[H]
\centering
\begin{tabular}{ccc}
 \hline
$\lambda_{C}$ & $z$ & $\nu$ \\ \hline
1.225 & $1.27 \pm 0.04$ & $0.95 \pm 0.05$ \\
1.805 & $1.23 \pm 0.03$ & $1.00 \pm 0.05$ \\
\hline 
\end{tabular}
\caption{Values of $z$ and $\nu$ obtained numerically for the topological transitions for $N_y=800$.}
\label{tab:zs_nus_discase}
\end{table}

\subsection{Fractal Analysis}
\begin{figure*}
	\centering
	%   \resizebox{0.8\textwidth}{!}{
	\includegraphics[width=\linewidth]{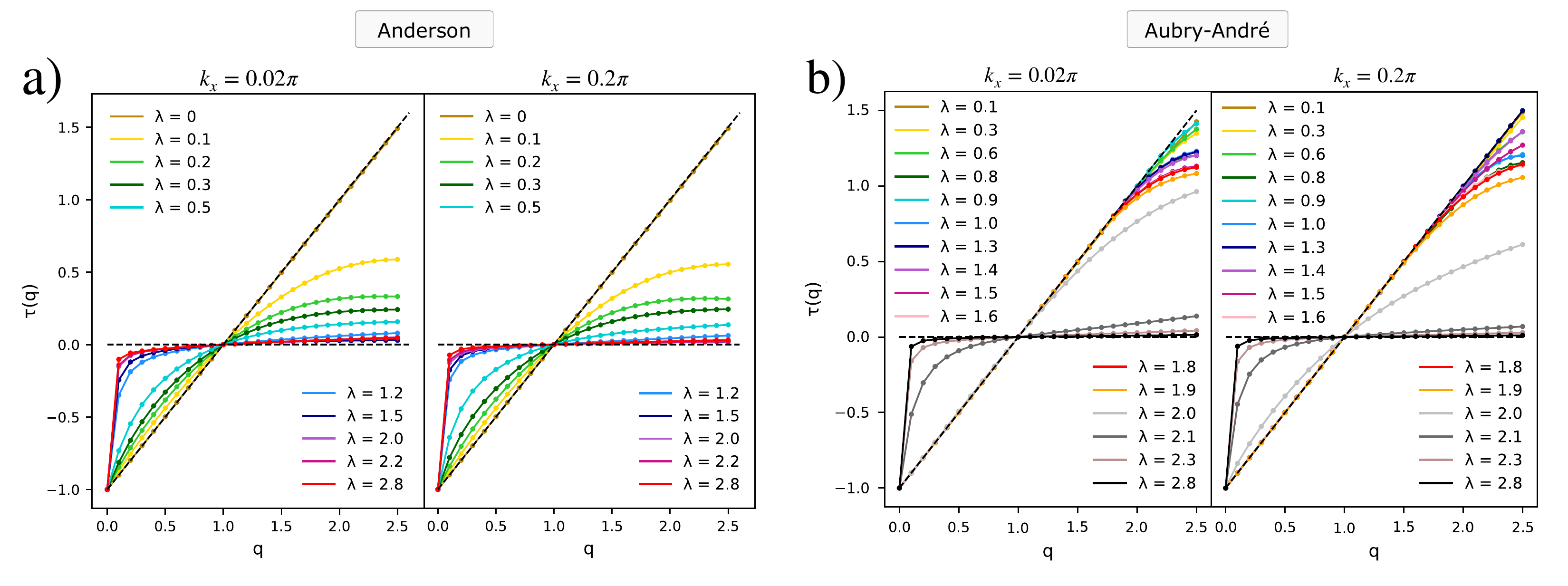}
	%    }
	\caption{Results of $\tau$ vs. $q$, for several values of disorder strength, $\lambda$, for $k_x=0.02\pi$ and $k_x=0.2\pi$, for a) Anderson disorder and b) Aubry-André disorder. In all cases, the IPR is averaged for the states within the energy range $E \in [0.05,1]$.}
	\label{fig:Multifractal}
\end{figure*}

\par One of the effects of Anderson transitions is the emergence of multifractality, which is characterized by fluctuations of eigenstates. These fluctuactions are manifested in the generalized inverse participation ratio. For a given eigenstate labeled by $m$, the generalized IPR is defined as:
\begin{equation}
    \text{IPR}(q)_{m}= \sum_{i}\left|\psi_{i}^{m}\right|^{2q},
    \label{eqn:disorder_pr_gipr}
\end{equation}
where, as before, $\psi_{i}^{m}$ is the wavefunction of the eigenstate $m$ at a site $i$.
At criticality, the generalized IPR behaves as \cite{wegner_1980}
\begin{equation}
    \text{IPR}(q) \sim L^{\tau(q)}
    \label{eqn:IPR_behaviour}
\end{equation}
where $L$ is the system size and the exponent $\tau(q)$ is defined in terms of a generalized dimension $D(q)$ as $\tau(q)=D(q)(q-1)$. In a metallic phase, $D(q)=d$ and for an insulating phase $D(q)=0$. Wavefunction multifractality is characterized by a $q$ dependent value of $D(q)$, whereas the cases of a constant $D(q)$ are single fractals \cite{evers_mirlin_2008}.
\par Here we want to make a simple fractal analysis of the system both for disorder and quasidisorder. 
We take $k_x$ at fixed values, such that system is reduced to an effective one dimension.
The IPR as a function of $q$ is calculated and averaged within the energy range $E \in [0.05,1]$. We fix the parameters $t=1$, $d=t/6$, $\mu=3d-4t$ and $B_y=0.5d$ and consider both the cases of Aubry-Andr\'e disorder and Anderson disorder. The following subintervals of $L$ are considered, to which a fit of an equation of the form of Eq. \ref{eqn:IPR_behaviour} is done:

\begin{equation}
\begin{split}
&L_1 = \\ &\{75, 100, 150, 175, 200, 255, 275, 400, 475, 600, 675, 800\}, \\
&L_2 = \{150, 175, 200, 255, 275, 400, 475, 600, 675, 800\}, \\
&L_3 = \{200, 255, 275, 400, 475, 600, 675, 800\}, \\
&L_4 = \{275, 400, 475, 600, 675, 800\}.
\end{split}
    \label{eqn:Intervalos}
\end{equation}
The obtained results are presented in Figs. \ref{fig:Multifractal} and \ref{fig:Multifractal2}.

\subsubsection{Anderson disorder}
\begin{figure*}
	\centering
	%   \resizebox{0.8\textwidth}{!}{
	\includegraphics[width=0.95\linewidth]{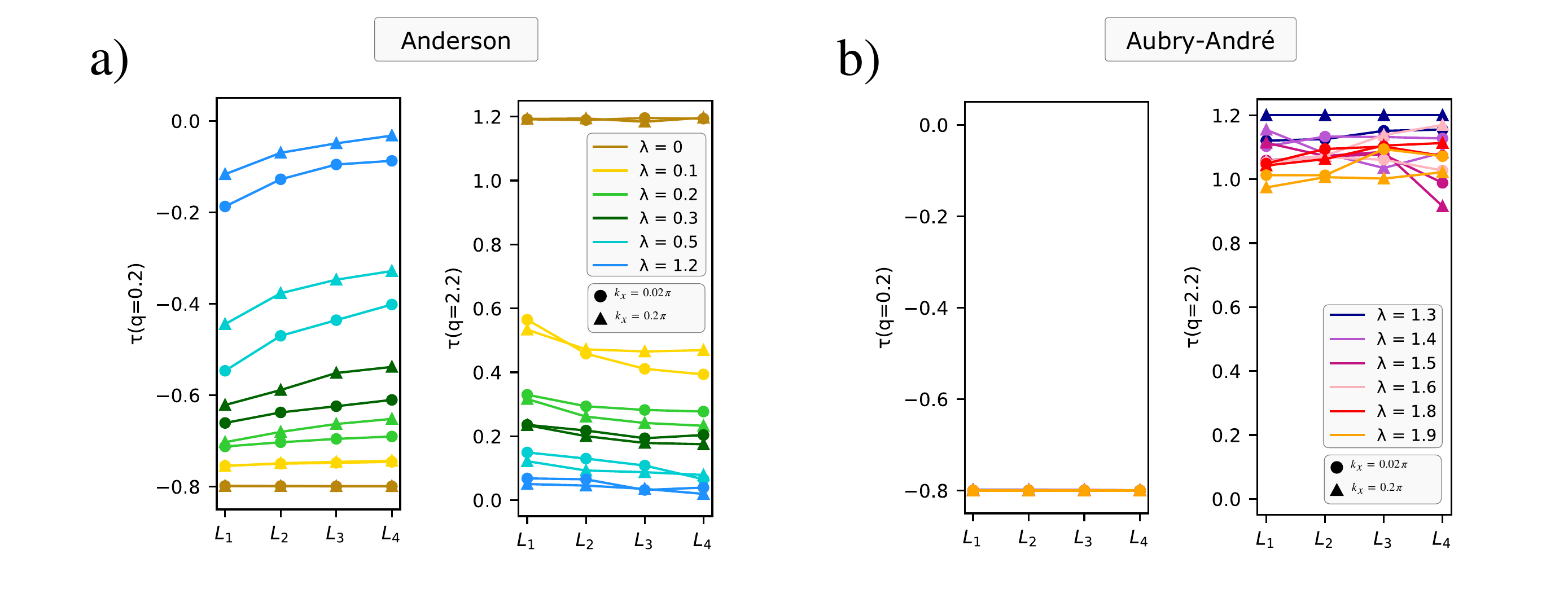}
	%    }
	\caption{a) Values of $\tau$ at different values of $q$ and Anderson disorder strength $\lambda$, for $k_x=0.02\pi$ and $k_x=0.2\pi$. b) Values of $\tau$ at different values of $q$ and quasidisorder strength $\lambda$, for $k_x=0.02\pi$ and $k_x=0.2\pi$. In all cases, the IPR is averaged for the states within the energy range $E \in [0.05,1]$.}
	\label{fig:Multifractal2}
\end{figure*}

Fig. \ref{fig:Multifractal}(a) shows the values of $\tau(q)$ for $k_x=0.02\pi$ and $k_x=0.2\pi$, for several values of $\lambda$ and considering the system size interval $L_1$. One thing that can be immediately noticed is that for the clean system, $\lambda=0$, the values of $\tau(q)$ closely follow the line $\tau(q)=(q-1)$, indicating that $D(q)$ is $q-$independent and equal to $1$. 
This is the expected behaviour of the clean system (taking a fixed $k_x$ where the system is reduced to one dimension) and reveals that the bulk states are extended in the $y$ direction.
For higher values of disorder, $\tau(q)$ approaches the line $\tau(q)=0$, where $D(q)=0$, suggesting the states are localized. For other values of disorder strength, starting at $\lambda=0.1$, $\tau(q)$ does not follow a behaviour characteristic either of $D(q)=1$ or $D(q)=0$. In order to take a conclusion, it is necessary to evaluate $\tau(q)$ as the system size tends to infinity.  To do this, the subintervals of $L$ in Eq. \ref{eqn:Intervalos} are considered, to which a fit of equation of the form of Eq. \ref{eqn:IPR_behaviour} is done. The results are presented in Fig. \ref{fig:Multifractal2}(a).
We find that for the values $\lambda=0.1$
and above, as larger values of $L$ are considered, the curves $\tau(q)$ approach $\tau(q)=0$. 
This confirms a localization of the bulk states in the thermodynamic limit for small values of disorder.
\subsubsection{Aubry-Andr\'e disorder}
Fig. \ref{fig:Multifractal2}(b) shows the values of $\tau(q)$ for $k_x=0.02\pi$ and $k_x=0.2\pi$, for several values of quasidisorder strength $\lambda$ for the size interval $L_1$. Unlike the previous case with Anderson disorder, we see that the results differ for each $k_x$, and that for some values of disorder strength $\tau(q)$ follows the line $q-1$ closely until some value of $q$ where the behaviour suddenly changes.  
In Fig. \ref{fig:Multifractal2}(b)  we show, as before, values of $\tau(q)$ fitted for the considered size intervals $L_1$, $L_2$, $L_3$ and $L_4$.
or lower values of $q$, $\tau$ remains at the values defined by the Eq. $\tau(q)=D(q)(q-1)$ with $D(q)=1$. However, at higher values of $q$, this behaviour changes. Contrary to the case with Anderson disorder, there is no clear tendency of $\tau(q)$ at increased system sizes, and the behaviour also depends on the value of $q$. This deviation from the $D(q)=1$ line is verified as soon as disorder is introduced, and suggests the system is in a multifractal regime. Accordingly, we see the appearance of critical bulk states in the system. 
From inspection of Fig. \ref{fig:Multifractal}(b) and of the corresponding values of $\tau(q)$ at larger system sizes, we identify a transition to a localized phase around $\lambda \in [2.0,2,1]$.

\section{\label{sec:conclusions}Conclusions}

\par In this work we studied a two-dimensional topological superconductor in the presence of a magnetic field. We introduced disorder and quasidisorder in the system with the aim of studying the effects on topological and localization properties.
Considering previous results on other systems such as insulators, semimetals and one-dimensional
superconductors, and the results we found on the effect of quasidisorder in two-dimensional superconductors,
we may expect that the results may be general considering gapless systems or topological systems (gapped or
gapless), in which regions displaying similar topological and localization properties may be found.

\par The system was first studied under a perpendicular magnetic field $B_z$. Four types of disorder were considered: Anderson disorder (two-dimensional), Anderson disorder (one-dimensional along $y$, uniform along $x$), Aubry-Andr\'e disorder (one-dimensional along $y$, uniform along $x$) and Aubry-Andr\'e disorder (two-dimensional). We observed that the response of the topological phases of the system differs depending on the type of disorder, and that quasidisorder induces topological phases in new regions of $B_z$, characterized by integer values of the Chern number $C$. The critical points at these phase boundaries were shown to become sharper as the system size increases, allowing us to conclude that the obtained phase diagrams apply to bigger system sizes.

\par The real space system was also briefly studied when a parallel magnetic field is applied in the $y$ direction. We studied the cases of bulk disorder, with the same four different spatial configurations. For two-dimensional Anderson disorder, we found that the average IPR of the system increases with an exponential behaviour as a function of $\lambda$ for $\lambda>1.5$. When uniformity in the $x$ direction is imposed in the Anderson disorder potential, we found that the IPR shows a linear increase as a function of $\lambda$, for $\lambda>0.5$. For Aubry-Andr\'e disorder, the behaviour of the average IPR of the system reveals the existence of two different regimes. In the first, the average IPR shows a slow increase with $\lambda$, and in the second the IPR greatly increases. The transition between the two regimes is located around $\lambda=2$. 
\par We then studied the system in a mixed ($k_x,y$) space with an applied parallel magnetic field. The clean superconducting system is known to possess flat bands in the gapless regime. At the corresponding values of $k_x$ these have a winding number $\mathcal{W}$ of 1, which is defined from reducing the two dimensional system to an effective one dimension. We showed that these are also characterized by a $\pi-$quantized Berry phase at the same values of $k_x$. 
\par We showed that the introduction of quasidisorder induces new gapless phases in
parameter regimes where they were absent in the clean case. For the $p$-wave system subject to a parallel magnetic field this leads to new regimes with Majorana flat bands. This is not only true for phases with a gapless bulk but also for gapped phases, where quasidisorder closes the bulk gap and Majorana flat bands appear. We then obtained the Berry phase with twisted boundary conditions and concluded the quasidisorder induced flat bands also have a quantized Berry phase of $\pi$.
For the noncentrosymmetric superconductor with added $s$-wave superconducting pairing and Rashba spin orbit coupling, we showed that new regimes with unidirectional Majorana edge states appear. 
In particular, we showed that for a phase where right-moving unidirectional edge states were present in the system, the introduction of quasidisorder leads to the appearance of edge modes in the opposite moving direction, and for a certain quasidisorder range these modes coexist in the system.

\begin{figure}
	\centering
	%   \resizebox{0.8\textwidth}{!}{
	\includegraphics[width=0.88\linewidth]{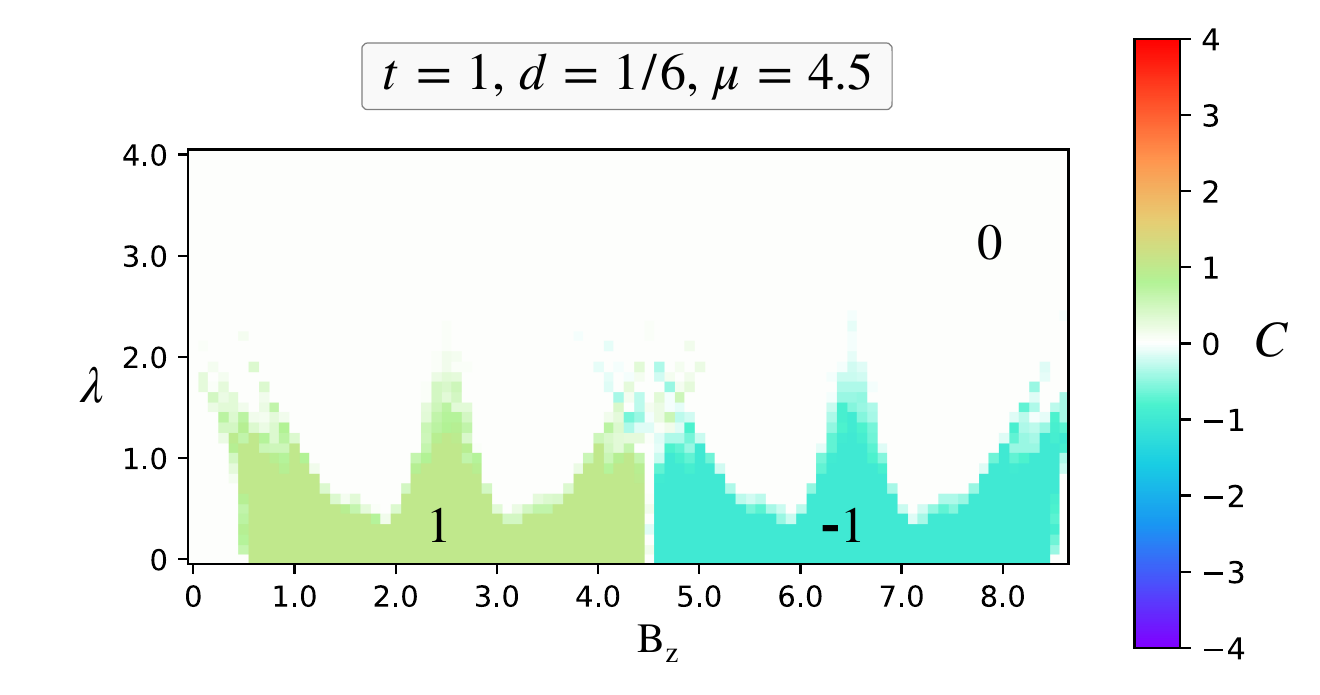}
	%    }
	\caption{Phase diagrams of a system with 20x20 sites indexed by the Chern number $C$, for several values of Aubry-Andr\'e disorder strength $\lambda$ and perpendicular magnetic field $B_z$, obtained for the average over 20 disorder configurations. The values of the parameters are $t=1$,  $d=1/6$ and $\mu=4.5$ .}
	\label{fig:chern_mu4_5}
\end{figure}
 
\begin{figure}
	\centering
	%   \resizebox{0.8\textwidth}{!}{
	\includegraphics[width=0.85\linewidth]{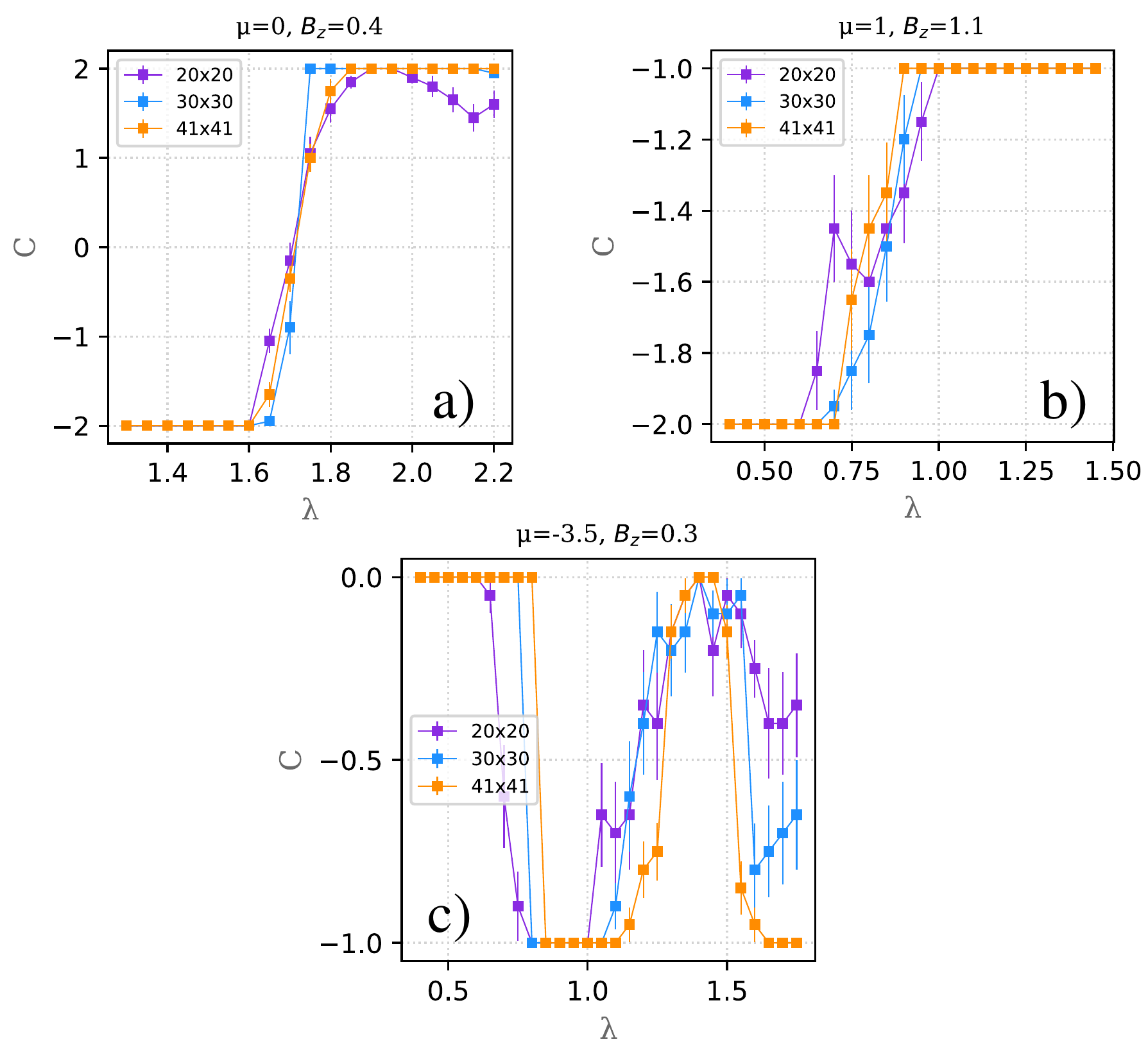}
	%    }
	\caption{Values of the Chern number $C$ vs. one-dimensional Aubry-Andr\'e disorder strength $\lambda$ for the system sizes $20\times 20$, $30 \times 30$ and $41 \times 41$ and for a) $t=1$, $\mu=0$, $d=0.6$, $B_z=0.4$, b) $\mu=1$, $d=0.6$, $B_z=1.1$, c) $\mu=-3.5$, $d=1/6$, $B_z=0.3$. The results were averaged over 20 disorder configurations.}
	\label{fig:scalingchern}
\end{figure}

\begin{figure}
	\centering
	%   \resizebox{0.8\textwidth}{!}{
	\includegraphics[width=0.85\linewidth]{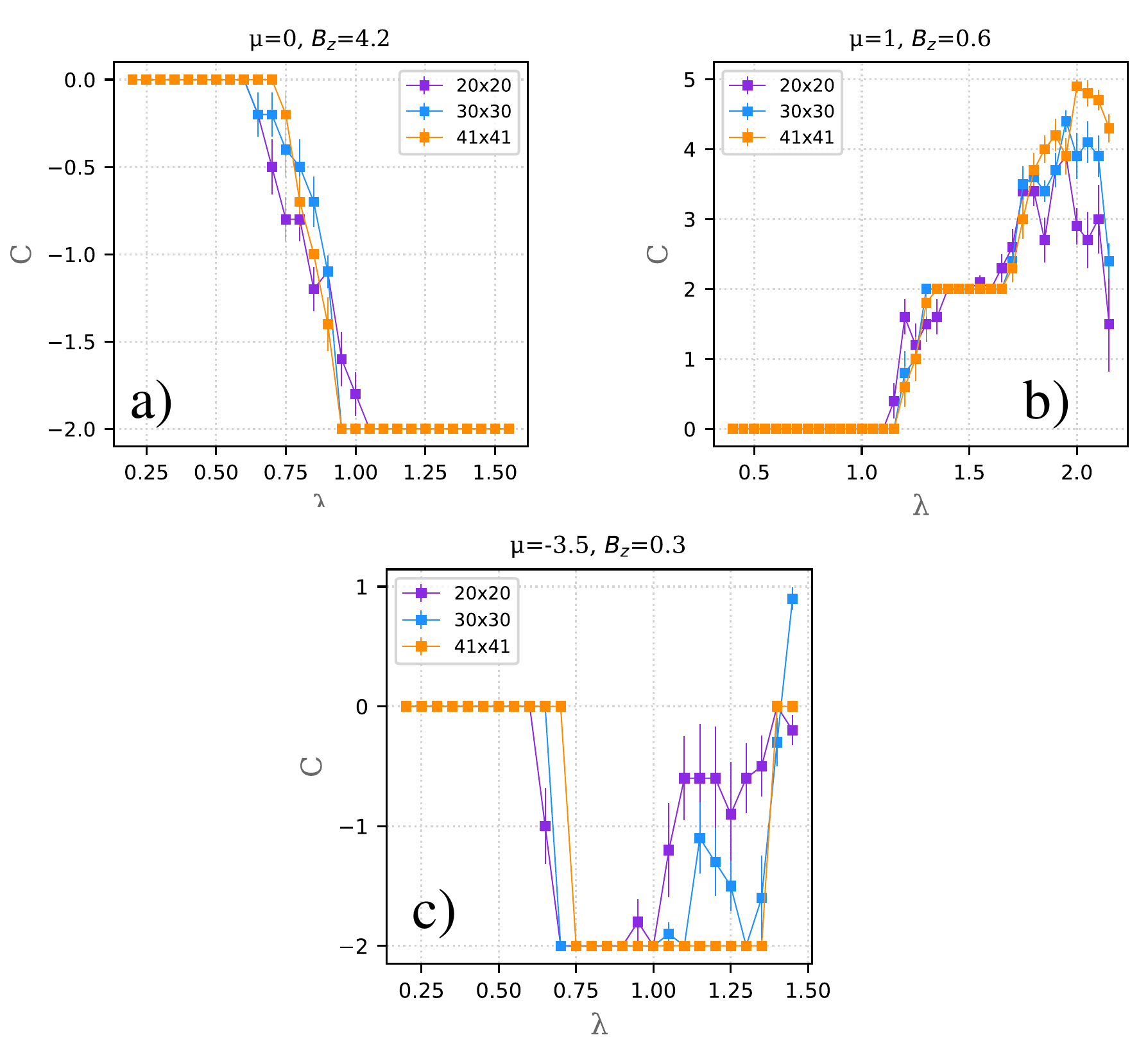}
	%    }
	\caption{Values of the Chern number $C$ vs. two-dimensional Aubry-Andr\'e disorder strength $\lambda$ for the system sizes $20\times 20$, $30 \times 30$ and $41 \times 41$ and for a) $t=1$, $\mu=0$, $d=0.6$, $B_z=0.4$, b) $\mu=1$, $d=0.6$, $B_z=1.1$, c) $\mu=-3.5$, $d=1/6$, $B_z=0.3$. The results were averaged over 10 disorder configurations.}
	\label{fig:scalingchern2}
\end{figure}

\par 
The identification of the quasidisorder induced topologically non-trivial flat bands 
with a quantized Berry phase of $\pi$ allowed us to study in detail two topological transitions, for the $p$-wave superconductor with a parallel applied magnetic field $B_y$. The two critical points were identified and studied by obtaining the density of induced Majorana bound states in relation to $k_x$ points. We found that the values of the critical points show almost no variation with the system size for systems bigger than $76$ sites along $y$.
The values of the dynamical critical exponents and correlation length critical exponents were obtained as
$z = 1.27 \pm 0.04$ and $\nu = 0.95 \pm 0.05$ for the first critical point and $z = 1.23 \pm 0.03$, $\nu = 1.00 \pm 0.05$ for the second critical point, which puts these transitions in novel universality classes. 
We then investigated the multifractal nature of the wavefunctions by calculating the values of $\tau(q)$ from the IPR values at several values of disorder, at the same parameter values as those in which the topological transition was studied. From the behaviour as the thermodynamic limit is approached, we concluded that the introduction of quasidisorder induces multifractality in the system. A transition to a localized regime was identified for $\lambda_{AA} \in [2.0,2.1]$. The same analysis was made for the system with Anderson disorder. The behaviour of $\tau(q)$ as the system size tends to infinity suggests that the introduction of Anderson disorder will drive the system to a localized phase (in the thermodynamic limit).

We have also shown that the average inverse participation ratio is not very sensitive to the magnitude of $B_z$ or $B_y$. Although $B_y$ leads to gapless behavior and $B_z$ in general leads to gapped behavior, and although each magnetic field direction leads to different topological properties and symmetries in the clean system, the localization properties are similar and the existence of critical states is also similar. 
It seems that a magnetic field in the $y$ direction, $B_y$, leads to a more localized behavior in the presence of a quasidisordered potential. For both magnetic field directions we found critical states and no mobility edges were found. We found a crossover as a function of $\lambda$ with a mixture of extended and critical states that grow in number as quasidisorder increases. 
In this context, Aubry-Andr\'e along $1d$ in the two dimensional system or along $2d$ does not lead to qualitatively different results (in the sense that the crossover in localization is seen for both cases), besides the differences in induced topology.

\begin{acknowledgments}
We acknowledge partial support from FCT through the Grant UID/CTM/04540/2019. M.F.M. acknowledges partial support through the grant (1801P.01102.1.01) QMSP2021 - CEFEMA - IST-ID.
\end{acknowledgments}

\appendix

\section{Additional details on the disordered model under a perpendicular magnetic field}\label{detailsChern}
\par In Figs. \ref{fig:ChernDiagrams}(h) and \ref{fig:ChernDiagrams}(i), we observe that for low magnetic fields the increase of quasidisorder induces topological phases in regions for which the Chern number was zero. In the clean system, however, these regions correspond to a phase that is topological and characterized by a finite value of $I(k_y)$. In 
Fig. \ref{fig:chern_mu4_5} we show a phase diagram for the parameters $d=1/6$ and $\mu=4.5$. In this case, the topological invariant $I(k_y=0,\pi)$ is trivial for low values of magnetic field ($B_z<0.5$) when $C=0$. Contrarily to what is observed in 
Fig. \ref{fig:ChernDiagrams}(i), there is no induced topological region with $C=-1$. This thus suggests that the reentrant regions observed in Fig. \ref{fig:ChernDiagrams}(h) and Fig. \ref{fig:ChernDiagrams}(i) can possibly be related with the topological nature of the phases characterized by $I(k_y)$ and with $C=0$.
\par To see how the different critical values for quasiperiodic disorder scale with the system size, three transitions for the phase diagrams obtained with Aubry-Andr\'e disorder along $y$ with uniformity along $x$ (third row in 
Fig. \ref{fig:ChernDiagrams}) and Aubry-Andr\'e in two dimensions (fourth row in Fig. \ref{fig:ChernDiagrams}) were considered, at fixed values of $B_z$ and $\mu$. The results are presented in Figs. \ref{fig:scalingchern} and \ref{fig:scalingchern2} for the system sizes $20\times 20$, $30 \times 30$ and $41 \times 41$. We found that within the considered system size range, the transitions become sharper as the size increases, thus suggesting that the obtained phase diagrams in Fig. \ref{fig:ChernDiagrams} should apply to larger systems.

\begin{figure}
\centering
\includegraphics[width=\columnwidth]{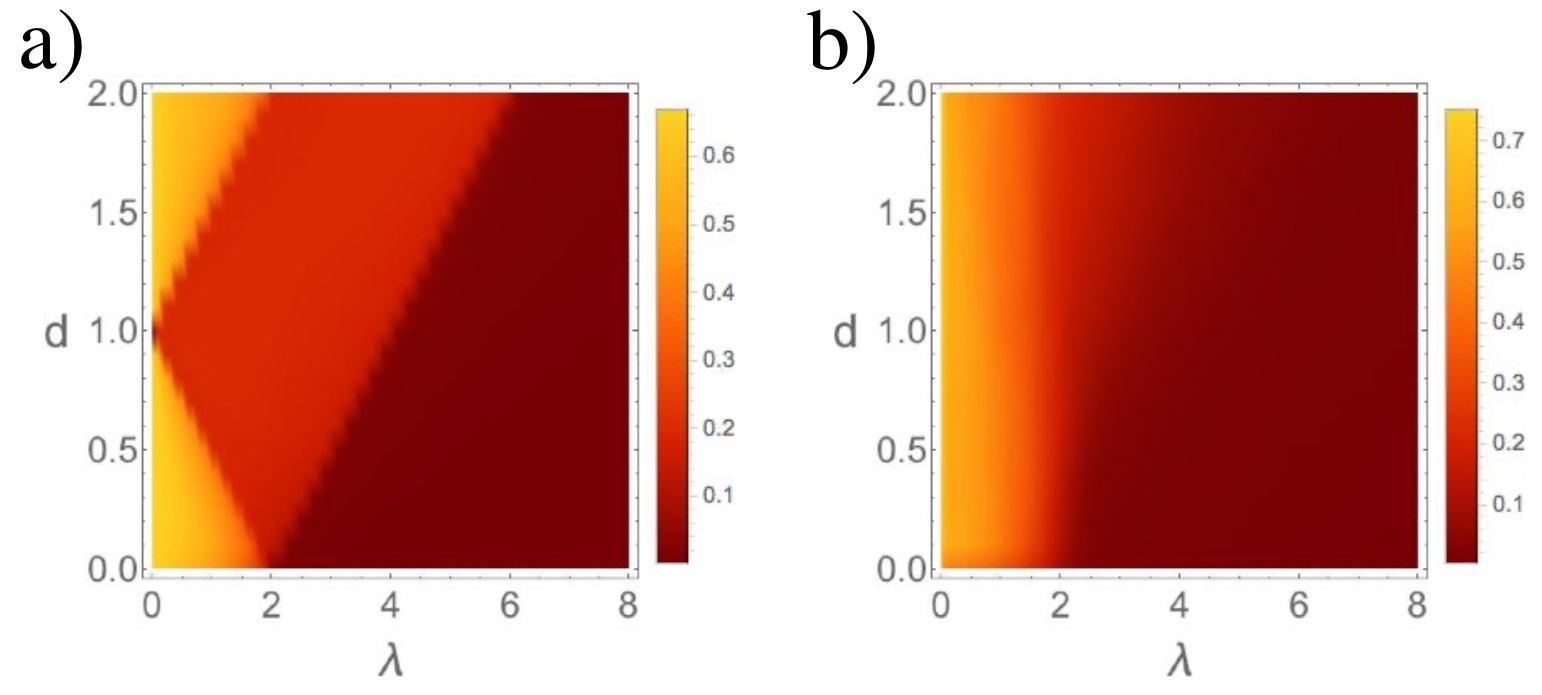}
\caption{\label{pr1}
Average participation ratio, APR, for a $p$-wave superconductor in the presence of Aubry-Andr\'e quasidisorder
for a) one-dimensional system and b) two-dimensional system, as a function of the superconducting pairing term $d$ and disorder strength $\lambda$.
}
\end{figure}

\begin{figure}
\centering
\includegraphics[width=0.9\columnwidth]{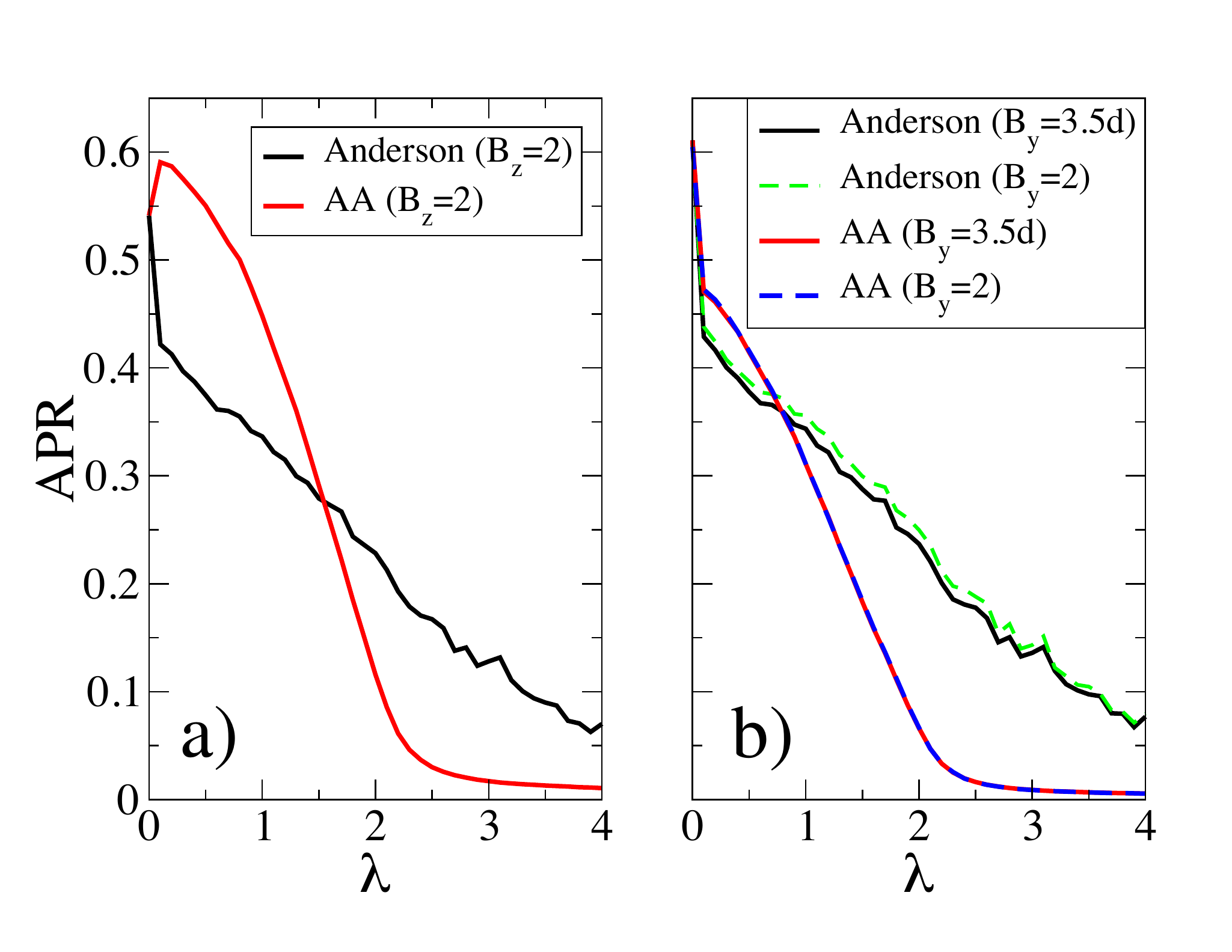}
\caption{\label{pr2}
Average participation ratio, APR, for a $2d$ system with Anderson or Aubry-Andr\'e (AA) quasidisorder,
for $d=t/6, \mu=-3.5$, in a a) perpendicular or b) parallel magnetic field.
}
\end{figure}

\section{Participation ratio of $2d$ Aubry-Andr\'e quasidisorder}\label{pr2d}

In Fig. \ref{fig:IPRS} we considered the average inverse participation ratio regarding both
disorder or quasidisorder along one spatial direction (the $y$ axis) and disorder or quasidisorder
in the plane, comparing a set of values of parallel magnetic field. In this Appendix we carry out a
more extensive analysis.
We want to focus our attention on the regime of increasing disorder strength, from small values to
larger values as localization takes place, in particular on the possible separation between extended,
critical and localized regimes. The inverse participation ratio is particularly useful to study the
transition to the localized regime, but is not as revealing in the extended-critical regimes. In this
Appendix we will consider instead the participation ratio, which is given by the inverse of Eq. \ref{eqn:disorder_pr_ipr}. It is of the order of one for extended
states, and becomes of the order of the inverse of the system size in the localized regime.

\begin{figure}
\centering
\includegraphics[width=0.85\columnwidth]{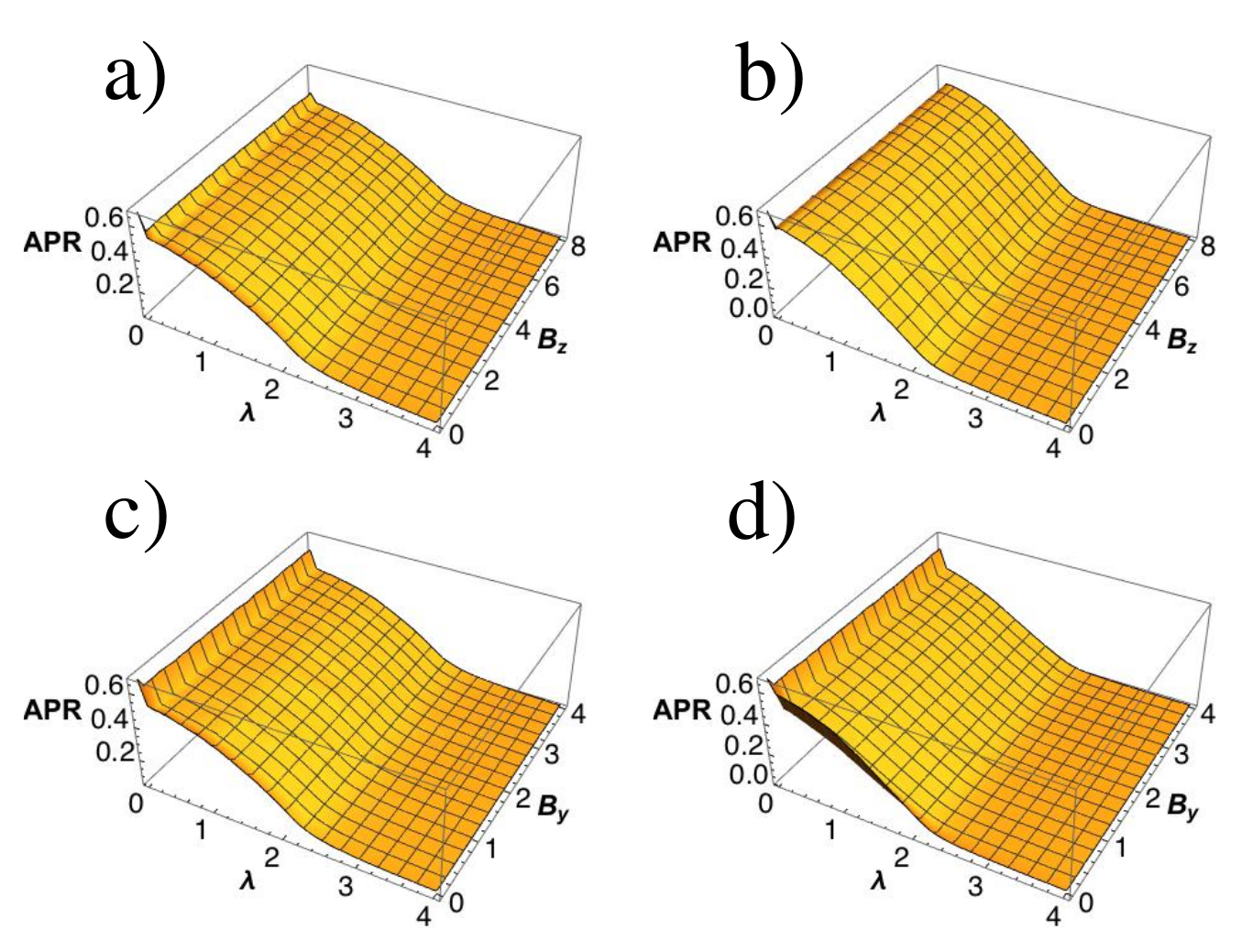}
\caption{\label{pr3}
Average participation ratio as a function of $\lambda$ for
perpendicular magnetic field, $B_z$, and a) $1d$ and b) $2d$ quasidisorder and
parallel magnetic field, $B_y$, and c) $1d$ and $2d$ quasidisorder. 
}
\end{figure}

\begin{figure}
\centering
\includegraphics[width=0.8\columnwidth]{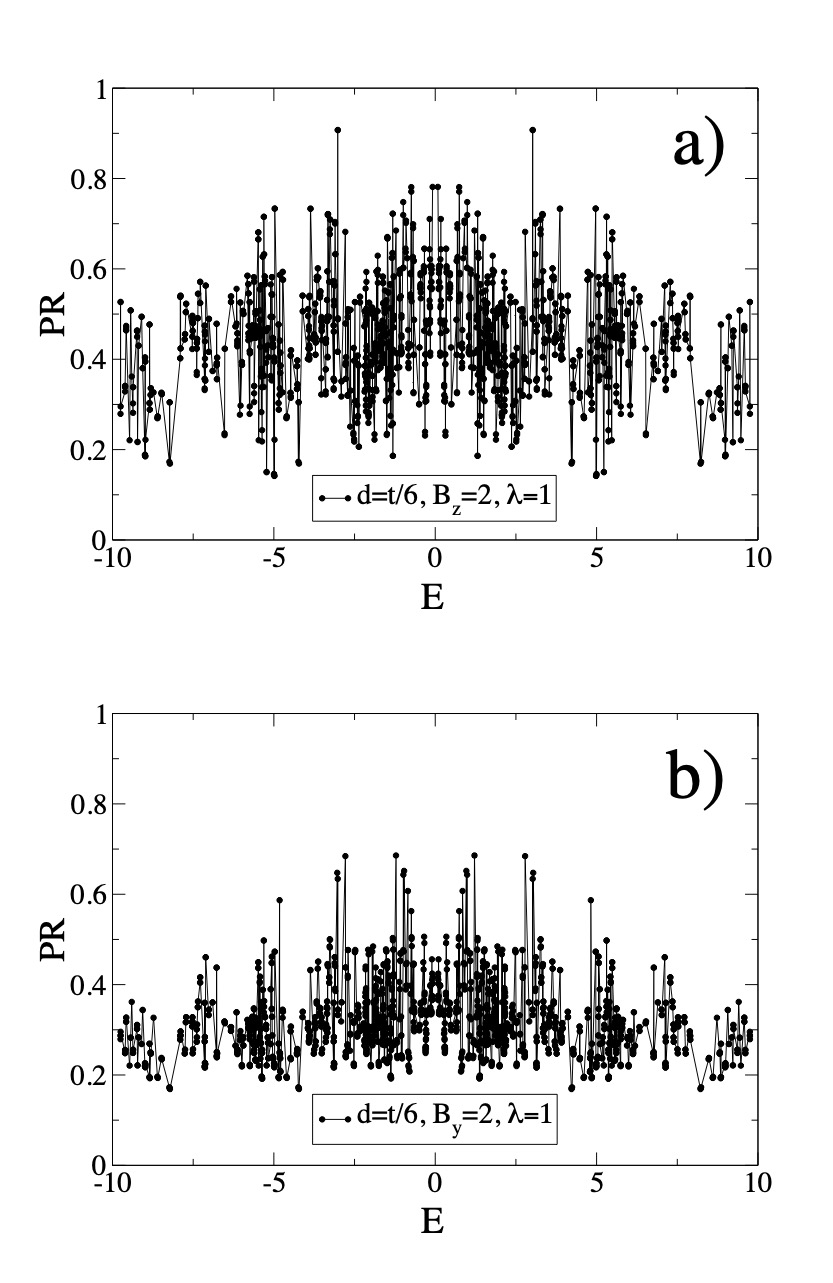}
\caption{\label{pr4}
Participation ratios of the eigenstates for
$\lambda=1, d=t/6, \mu=-3.5$, as a function of energy for a) $B_z=2$ and b) $B_y=2$.
}
\end{figure}

In the case of a one-dimensional $p$-wave superconductor (Kitaev model) in the presence of an Aubry-Andr\'e
potential and with no magnetic field, 
it is known that the three regimes of extended, critical and localized states are present, as one
changes the amplitude of the pairing, $d$, and the quasidisorder strength, $\lambda$ \cite{kitaevaa}.
The average participation ratio (APR), considering a single disorder configuration, 
is shown in Fig. \ref{pr1}(a), where clear transitions are shown separating
the various regimes (here we only consider a disorder configuration, but the results
are characteristic of a larger set of disorder configurations). 
The average participation ratio has different plateaus as the parameters change.
Considering a two-dimensional $p$-wave superconductor with two-dimensional quasidisorder, there is no clear
transition between the regimes and one finds crossovers as $\lambda$ increases. The system sizes considered
in Fig. \ref{pr1}(b) are small ($21 \times 21$), but a smooth transition showing the decreasing of the
average participation ratio seems to indicate no clear separation of a critical regime before the localized
phase takes place.

In Fig. \ref{pr2} we consider the average participation ratio for the two-dimensional case in the presence
of a perpendicular or parallel magnetic field, for the cases of Anderson disorder and quasidisorder.
These results highlight the extended/critical regimes at lower values of $\lambda$. 
Anderson disorder behaves similarly for the two magnetic field directions, and quasidisorder leads to higher
values of the average participation ratio as disorder increases, with a sharper transition to the localized regime,
as shown in Fig. \ref{fig:IPRS}. In the case of the parallel magnetic field quasidisorder has a stronger localization effect.
A difference with respect to the perpendicular magnetic field is the existence of gapless states, more sensitive
to the long-range disorder associated with the Aubry-Andr\'e potential. In the case of perpendicular
magnetic field, the system remains gapped (except at the transitions between the various topological
regimes) and therefore is expected to be less sensitive to the quasidisorder potential.

The crossover behavior is clearly seen in Fig. \ref{pr3}, independently of the magnetic field
direction ($B_z$ or $B_y$). 
Also, the consideration of quasidisorder along the $y$ direction and periodic along $x$ or quasidisorder that is fully two-dimensional leads to similar
results.
Some differences are visible for small magnetic fields or small values of $\lambda$.
Except for these regions, the average participation ratio is quite independent of the amplitude
of the magnetic field, but the effect of a parallel magnetic field is more significant, as
discussed.

\begin{figure}
\centering
\includegraphics[width=0.95\columnwidth]{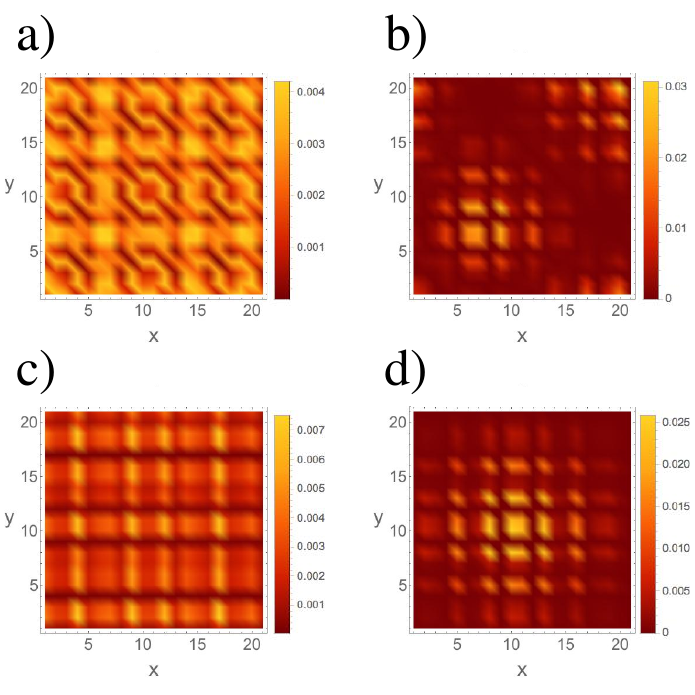}
\caption{\label{pr5}
Wave function amplitudes for the parameters of Fig. \ref{pr4}, as a function of space location in a
$21 \times 21$ system, for states with a) $B_z=2$ and participation ratio $PR=0.781$,  
b) $B_z=2$ and $PR=0.186$, c) $B_y=2$ and $PR=0.643$, d) $B_y=2$ and $PR=0.169$. 
}
\end{figure}

In order to have a better understanding of the possible existence of critical
states in the regime prior to the transition to localization, we show a typical case
in Fig. \ref{pr4} of the participation ratios of the various eigenstates, as a function
of their energies, for perpendicular and parallel magnetic fields. The parameters are chosen
so that we are in an intermediate regime, where the average participation ratio is in the
crossover between fully extended and localized states. There are significant fluctuations
between states with high participation ratios (characteristic of extended states) and
low participation ratios (characteristic of intermediate, critical, states) but still
larger than values that correspond to the localized regime. The results do not show a
mobility edge, and the states mix throughout the energy range. Also, as $\lambda$ increases,
we have found that the percentage of critical-like states increases, explaining the crossover
behavior. The extended versus critical character nature of the states is shown in Fig. \ref{pr5},
where a few wave functions are shown (for the system with periodic boundary conditions in both directions), characteristic of extended and critical states coexisting in the same energy spectrum.

\section{Energy spectra evolution for the noncentrosymmetric superconductor with Aubry-Andr\'e and Anderson disorder in $(k_x,y)$ space}\label{sec:energyspectra}

\begin{figure}
	\centering
	%   \resizebox{0.8\textwidth}{!}{
	\includegraphics[width=\linewidth]{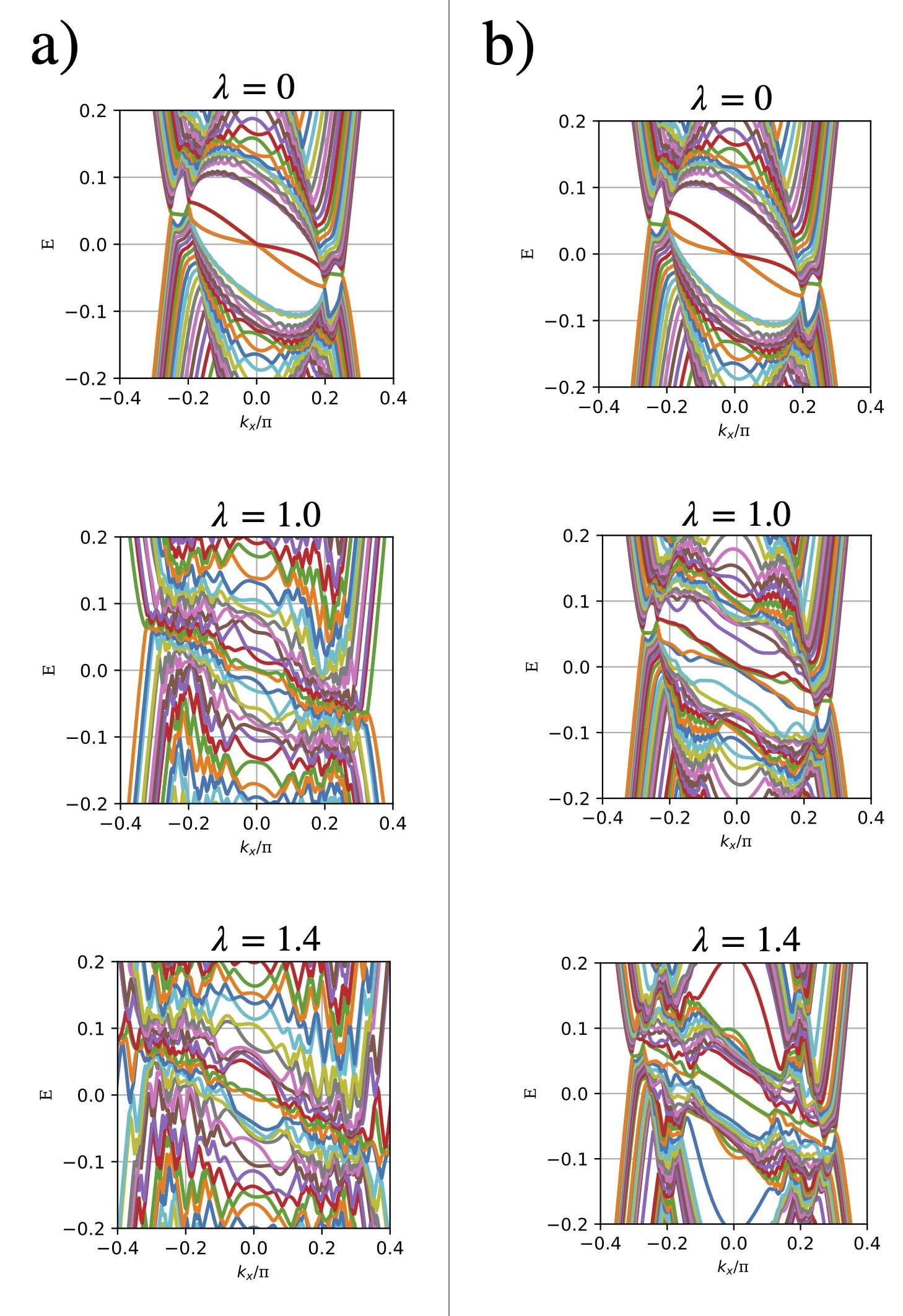}
	%    }
	\caption{Energy spectra evolution with  a) Anderson disorder 
and b) Aubry-Andr\'e disorder, for $B_y=d$, $\alpha=0.2d$ and $\Delta_s =0.5d$.}
	\label{fig:Byd_NCS}
\end{figure}

\par Here we present results for the evolution of the energy spectra for the noncentrosymmetric superconductor, with spin orbit coupling $\alpha$ and mixed $p$ and $s$-wave pairings, subject to Anderson and Aubry-Andr\'e disorder in the $(k_x,y)$ space. We take the same parameter values for $t$, $d$ and $\mu$ as in section \ref{sec:disorderedperparallel_1}, and consider two different cases: $B_y=d$, $\alpha=0.2d$, $\Delta_s =0.5d$ and $B_y=4d$, $\alpha=0.2d$ and $\Delta_s =0.3d$. 

%\begin{figure}[h]

\par
\begin{figure*}
	\centering
	%   \resizebox{0.8\textwidth}{!}{
	\includegraphics[width=0.9\linewidth]{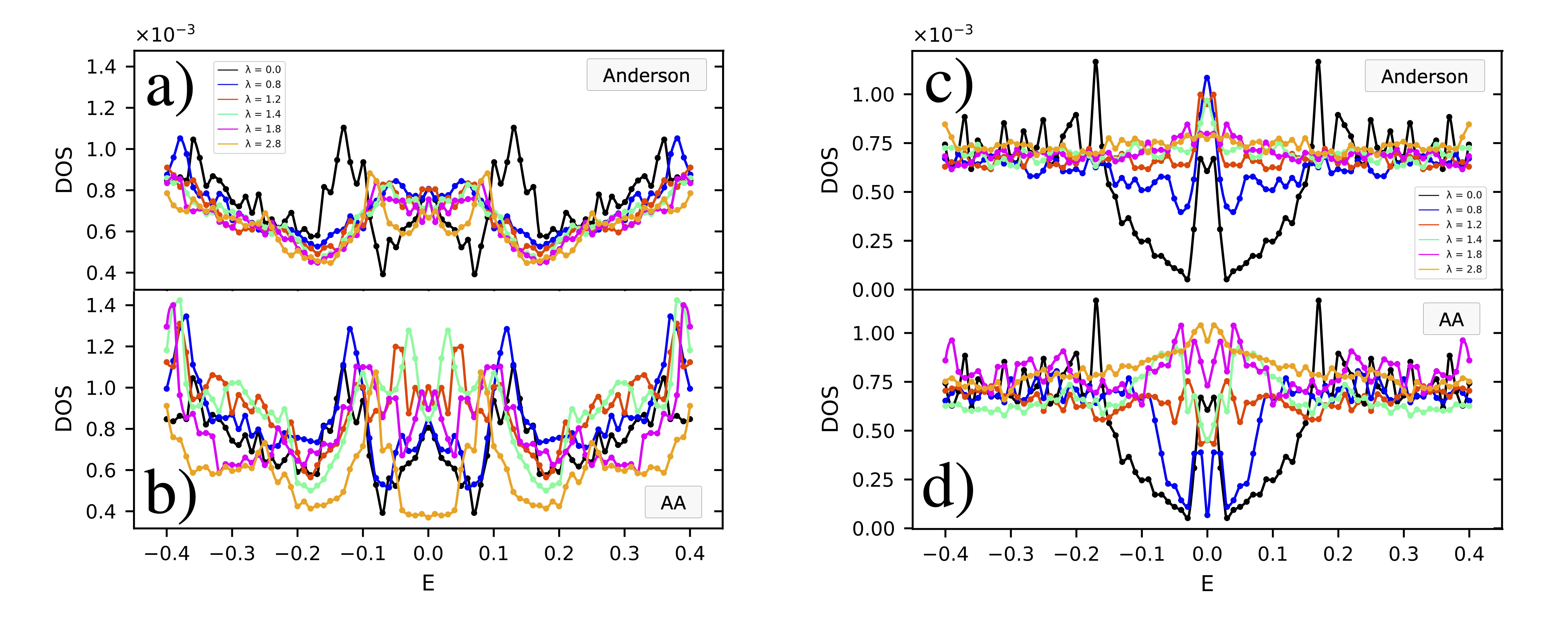}
	%    }
	\caption{
Density of states evolution with  a) Anderson disorder 
and b) Aubry-Andr\'e (AA) disorder, for $B_y=d$, $\alpha=0.2d$ and $\Delta_s =0.5d$ and with c) Anderson disorder 
and d) Aubry-Andr\'e disorder for $B_y=4d$, $\alpha=0.2d$, and $\Delta_s = 0.3d$.}
	\label{fig:DOS_NCS}
\end{figure*}

 In Fig. \ref{fig:Byd_NCS} we present the evolution of the energy spectrum for a) Anderson disorder and b) Aubry-Andr\'e disorder, for $B_y=d$, $\alpha=0.2d$, $\Delta_s =0.5d$. As Anderson disorder is introduced in the system, the edge states are destroyed and the considered energy range gets filled with bulk states, but the tilt of the spectrum is preserved. As a result, the flat bands which were previously lifted due to the introduction of finite values of $\alpha$ and $\Delta_s$ do not collapse to zero energy. For high values of $\lambda$ the density of states exhibits two peaks which result from the inclination of the bulk energy spectrum (Fig. \ref{fig:DOS_NCS} a) ). 
\begin{figure}
	\centering
	%   \resizebox{0.8\textwidth}{!}{
	\includegraphics[width=\linewidth]{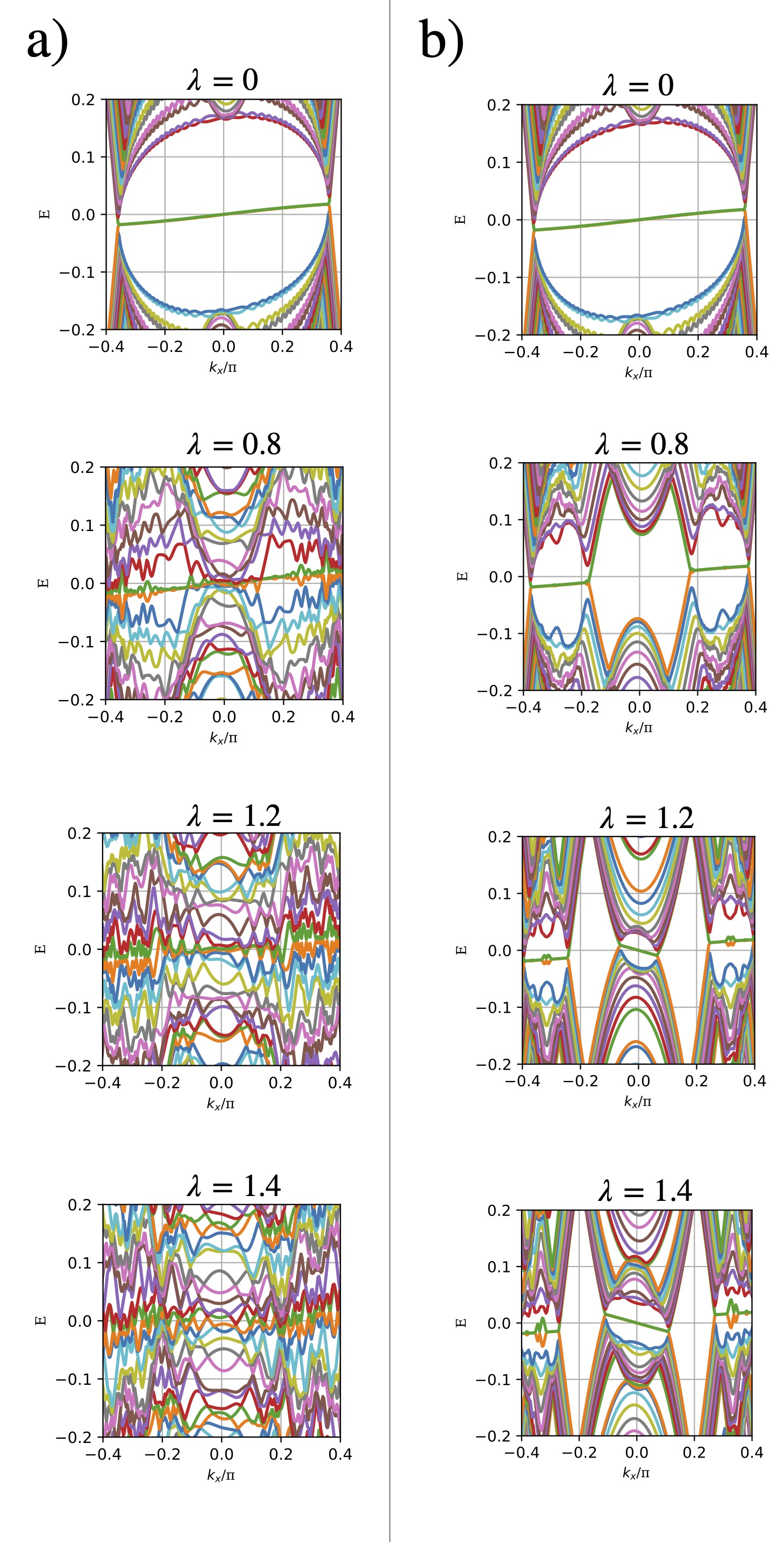}
	%    }
	\caption{Energy spectra evolution with  a) Anderson disorder and b) Aubry-Andr\'e disorder, for $B_y=4d$, $\alpha=0.2d$, and $\Delta_s = 0.3d$.}
	\label{fig:Byd_MES}
\end{figure}
\par In Fig. \ref{fig:Byd_NCS}(b), as Aubry-Andr\'e disorder is introduced, we see an evolution that is similar 
to Fig. \ref{fig:Evolution_Byd}(b), but instead of new flat band regimes, new unidirectional edge states appear. Unlike what happens for Anderson disorder, at high values of $\lambda$ a gap opens for values of $k_x$ around $k_x=0$ (although the bulk as a whole remains gapless). This is reflected in the density of states, that drops around $E=0$ for higher disorder values. Similarly to what was observed for Anderson disorder, the tilt of the energy spectrum is preserved as disorder increases.

\par In Fig. \ref{fig:Byd_MES}, the clean system is in the regime where unidirectional MESs appear. The values of the $p$-wave pairing and spin orbit term are kept constant in relation to the case of Fig. \ref{fig:Byd_NCS}, but the $s$-wave pairing term is decreased from $\Delta_s=0.5d$ to $\Delta_s=0.3d$ and the 
magnetic field is increased from $B_y=d$ to $B_y=4d$. The spectrum acquires a tilt in the opposite direction if compared to the clean system in Fig. \ref{fig:Byd_NCS}, as a result of the increased magnetic field. With Anderson disorder, 
Fig. \ref{fig:Byd_MES}(a), The unidirectional Majorana edge states are robust to small values of disorder strength but as disorder increases the structure of the band is lost, as bulk states fill the lower energy values. This differs from 
Fig. \ref{fig:Byd_NCS}(a) where the tilt of the spectrum is preserved even at higher values of disorder. As disorder is increased, there is at first an increase in the value of the DOS at zero energy, which then decreases for higher values of disorder. For $\lambda>1.8$ the density of states becomes nearly constant in the considered range of $E \in [-0.4,0.4]$.
\par In Fig. \ref{fig:Byd_MES}(b), when a certain value of Aubry-Andr\'e disorder is reached, "flipped" unidirectional states appear in the system. This is seen clearly in  Fig. \ref{fig:Byd_MES}(b) for the values of $\lambda=1.2$ and $\lambda=1.4$, as a band with negative slope appears for values of $k_x$ around $k_x=0$. At $\lambda=1.2$ there is a coexistence of unidirectional "flipped" left-moving edge modes (with negative slope) around $k_x=0$ and right-moving edge modes (with positive slope) for higher (absolute) values of $k_x$. A backflow current that balances the current on the edges is created on the bulk: extra right or left moving modes will appear depending on the net current on the edges.

%\newpage
%\section{Density of states evolution with Aubry-Andr\'e and Anderson disorder in $(k_x,y)$ space}
%CHANGE THIS FIGURE
%In Fig. \ref{fig:DOS} the evolution of the density of states is presented. Figs. \ref{fig:DOS}(a), (b), (e) and (f) are discussed in section \ref{sec:disorderedperparallel_1} and Figs. \ref{fig:DOS}(c), (d), (g) and (h) are discussed in the appendix section \ref{sec:energyspectra}.

%\newpage
%


\begin{thebibliography}{70}
\bibitem{sizes}
Here we limit ourselves to the analysis of a system of size $41\times41$, although draft results for a system of size $20\times20$ followed the same behaviour as the one discussed.
\end{thebibliography}
\end{document}